\documentclass[aps,pre,twocolumn,amsmath,amssymb,amsfonts]{revtex4}
\usepackage{epsfig}
\usepackage{graphicx}
\usepackage{dcolumn}
\usepackage{bm}
\usepackage{epsfig}
\usepackage{graphicx}
\usepackage{dcolumn}\usepackage{braket}
\usepackage{color}
\usepackage{ulem}
\usepackage{bm}
\usepackage{soul, xcolor}
\usepackage{xcolor,cancel}
\usepackage[titletoc,title]{appendix}

\begin{document}
\def\red{\textcolor{red}}
\setstcolor{red}
\definecolor{ao}{rgb}{0.0, 0.0, 1.0}
\definecolor{em}{rgb}{0.25, 0.0, 1.0}
\definecolor{eb}{rgb}{0.06, 0.2, 0.65}

\author{Itzhak Fouxon$^{1,2}$}\email{itzhak8@gmail.com}
\author{Joshua Feinberg$^{2}$}\email{joshua@ph.technion.ac.il}
%\author{Petri K\"{a}pyl\"{a}$^{3,4,5,6,7}$}\email{pkaepyl@uni-goettingen.de}
%PJK: reduced affiliations
\author{Petri K\"{a}pyl\"{a}$^{3}$}\email{pkaepyl@uni-goettingen.de}
\author{Michael Mond$^{1}$}\email{mondmichael@gmail.com}
\affiliation{$^1$ Department of Mechanical Engineering, Ben-Gurion University of the Negev, Beer Sheva 84105, Israel}
\affiliation{$^2$ Department of Mathematics and Haifa Research Center for Theoretical Physics and Astrophysics, University of Haifa,
Haifa 31905, Israel}
\affiliation{$^3$ Georg-August-Universit\"{a}t G\"{o}ttingen, Institut f\"{u}r Astrophysik, G\"{o}ttingen, Germany}
%\affiliation{$^4$ Leibniz-Institut f\"{u}r Astrophysik, Potsdam, Germany}
%\affiliation{$^5$ Department of Computer Science, ReSoLVE Centre of Excellence, Aalto, Finland}
%\affiliation{$^6$ Max-Planck-Institut f\"{u}r Sonnensystemforschung, G\"{o}ttingen, Germany}
%\affiliation{$^7$ NORDITA, KTH Royal Institute of Technology and Stockholm University, Stockholm, Sweden}

\begin{abstract}

The Navier-Stokes equations generate an infinite set of generalized
Lyapunov exponents defined by different ways of measuring the distance
between exponentially diverging perturbed and unperturbed
solutions. This set is demonstrated to be similar, yet different, from
the generalized Lyapunov exponent that provides moments of distance
between two fluid particles below the Kolmogorov scale. We derive
rigorous upper bounds on dimensionless Lyapunov exponent of the fluid
particles that demonstrate the exponent's decay with Reynolds number
$Re$ in accord with previous studies. In contrast, terms of cumulant
series for exponents of the moments have power-law growth with
$Re$. We demonstrate as an application that the growth of small
fluctuations of magnetic field in ideal conducting turbulence is
hyper-intermittent, being exponential in both time and Reynolds
number. We resolve the existing contradiction between the theory, that predicts slow decrease of dimensionless
Lyapunov exponent of turbulence with $Re$, and observations exhibiting quite fast growth. We demonstrate
that it is highly plausible that a pointwise limit for
the growth of small perturbations of the Navier-Stokes equations exists.

\end{abstract}
\title{Reynolds number dependence of Lyapunov exponents of turbulence and fluid particles}
\maketitle

\section{Introduction}

The Navier-Stokes equations are an infinite-dimensional dynamical system where small perturbations of its solutions grow exponentially. This results in a finite time over which the evolution can be predicted. Thus for instance thermal fluctuations change the macroscopic turbulent flow quite quickly \cite{ruelle7}, the fact that underlies the mechanism via which changes of motion of a single electron can cause global atmospheric changes in a couple of weeks \cite{ch}.

The growth of small perturbations is traditionally described by a
Lyapunov exponent \cite{oseledets,ruelle,ruelle7}. The exponent
provides the logarithmic growth rate of the distance between the
perturbed and unperturbed solutions. The definition of the distance
involves introducing a norm in the functional space and it is not
obvious which norm must be used. Thus it is usual to assume that the
theorem on the existence and realization-independence of the Lyapunov
exponent $\lambda^v$, which was proved for finite-dimensional systems
\cite{oseledets}, generalizes to the Navier-Stokes equations
\cite{mohan,boffetta,pnrom}. This would tell that
$\lim_{t\to\infty}t^{-1}\ln (||\delta\bm v(t)||/||\delta\bm v(0)||)$
is independent of the initial conditions on the perturbation flow
$\delta\bm v(t)$ and also of the unperturbed flow. In the case of a
finite-dimensional system any definition of the norm $||\delta\bm
v(t)||$ would result in the same limit. This however is not
necessarily the case for the infinite-dimensional systems where
different definitions of the norm can produce different limits. We
demonstrate here that for the Navier-Stokes equations
$\gamma^v(p)\equiv \lim_{t\to\infty}t^{-1}\ln (||\delta\bm
v(t)||_p/||\delta\bm v(0)||_p)$, with the $L_p-$norms $||\delta\bm
v(t)||_p \equiv \left(\int |\delta v|^p d\bm x \right)^{1/p}$, differ
for different $p$ by powers of the Reynolds number $Re$. Therefore the
usually used definition with the $L_2-$norm
\cite{mohan,boffetta,pnrom}, giving $\lambda^v=\gamma^v(2)$, leaves
%PJK: This is confusing because no instability has been mentioned so far.
%PJK: Or do you mean the divergence of the solutions?
outside many essential details of the divergence of the solutions.

There is a controversy in the current knowledge which we propose to be resolved by the different $Re-$dependence of $\gamma^v(p)$ for different $p$. The theory of \cite{cr} predicts that the dimensionless Lyapunov exponent, obtained by multiplying $\lambda^v$ with the Kolmogorov time-scale \cite{frisch}, decays with $Re$ as a power-law with small exponent. However \cite{mohan} observed in direct numerical simulations a power-law growth with an appreciable exponent.

We explain the reason for the discrepancy. The study of $\lambda^v$ performed in \cite{cr} relies on Ruelle's assumption \cite{ruelle7} that the exponent can be estimated as the average of the inverse of the minimal time-scale of turbulence. This time-scale is given by the local viscous time-scale of the flow $t_{\nu}(\bm x)$, see e.g. \cite{frisch}. Roughly the assumption is rooted in the observation that perturbations grow due to local stretching of fluid elements whose rate is determined by the local velocity gradients given by $\sim t_{\nu}^{-1}(\bm x)$. Thus \cite{cr} assumed $\lambda^v=\langle t_{\nu}^{-1}\rangle$ where angular brackets stand for averaging. In fact, there is no ground to the last equality because local growth rates fluctuate strongly and it is by no means evident which of them determines the global growth rate of the perturbation $\lambda^v$. Indeed, intermittency of turbulence implies that $t_{\nu}(\bm x)$ undergoes strong spatial fluctuations with amplitude proportional to powers of $Re$, as seen e.g. from the multifractal model \cite{frisch}. The local growth at the rate given by $t_{\nu}^{-1}(\bm x)$ implies $||\delta\bm v(t)||_p\sim\left(\int \exp(pt/t_{\nu}(\bm x))d\bm x \right)^{1/p}$ which reduces to the space average $\langle t_{\nu}^{-1}\rangle$ only at $p\to 0$ (where $||\delta\bm v(t)||_p\sim\left(1+pt\langle t_{\nu}^{-1}\rangle\right)^{1/p}\sim \exp(t\langle t_{\nu}^{-1}\rangle)$ gives $\gamma^v(p=0)=\langle t_{\nu}^{-1}\rangle$. We remark that technically $L_p$ for $0<p<1$ is only a quasi-norm however this is irrelevant here and below). It is evident that at $p=2$, used for calculating $\lambda^v$, the averaged quantity is very different from $t_{\nu}^{-1}$. This results in a power of Reynolds number difference between $p=2$ and $p\to 0$ cases explaining the observations of \cite{mohan}. The above was irrelevant for the Ruelle's work \cite{ruelle7} whose purpose was making estimates within the Kolmogorov theory that disregards intermittency.

%However the predictions obtained in this way are not valid for the Navier-Stokes turbulence which is strongly intermittent. The reason are correlations between the local growth rate and the local values of the growing perturbations. For instance these correlations cause significant difference between the average local logarithmic growth rate of the perturbation and the average logarithmic growth rate of its energy.

Dependence of the limit for the Lyapunov exponent on the norm's definition is in sharp contrast with finite-dimensional dynamical systems. There any definition of the distance between the solutions produces the same exponent. Indeed, if an $l_p-$norm of a finite dimensional vector grows exponentially, then the growth exponent equals the maximal growth exponent of the components. This implies that the growth exponents of all $l_p$ norms are the same. This is not so when the number of degrees of freedom is infinite because the $L_p$ norm depends not only on the local value of the field but also on the space fraction where the value holds, that also depends on time exponentially. As a result, infinite dimensionality of the Navier-Stokes equations demands reconsideration of the facts established for Lyapunov exponents of finite-dimensional systems.

The above considerations raise the issue of whether the limits for the Lyapunov exponent are realization-independent irrespective of which norm is used. If they are, which seems to be the case, then which of them and when must be used. The usage of different $p-$norms would lead to a power of $Re$ difference of the corresponding Lyapunov exponents $\gamma^v(p)$, which can be crucial in astrophysical or other applications with very high $Re$.

The rationale for using $\lambda^v$ defined by the $L_2$ norm is that it describes the difference of energies of perturbed and unperturbed solutions (Caution must be exerted though. The energy difference is the sum of $||\delta\bm v||_2^2$ which is quadratic in the perturbation and a non-trivial term which is linear in the perturbation and could dominate the difference. Thus $||\delta\bm v||_2^2$ could be considered as the lower bound for the energy difference). However the growth is intermittent and it is possible to have a large energy difference when the local difference of perturbed and unperturbed flows $\delta\bm v(t, \bm x)$ is still small in most of the space. Thus our work indicates that the major fraction of space is described by an exponent very different from $\lambda^v$. We demonstrate that it is plausible that the perturbation grows exponentially at almost every spatial point with asymptotically the same growth exponent. This exponent is readily seen to be given by the derivative of $\gamma^v(p)$ at $p=0$ and differs from $\lambda^v$ by a power of $Re$. The difference holds because the asymptotic pointwise growth rate is attained very non-uniformly in space.

Another norm of interest is $L_{\infty}$ with $\gamma^v(p=\infty)$ providing the growth exponent of the maximal value of the perturbation. In fact, the intermittency of turbulence causes strong inhomogeneity of the perturbation growth. It is characterized by bursts and can only be described by using the infinite set of Lyapunov exponents as given by the full function $\gamma^v(p)$.

The infinite set of Lyapunov exponents $\gamma^v(p)$ is qualitatively similar to another set of Lyapunov exponents associated with the turbulent flow. These exponents also describe the exponential growth of the distance between two infinitesimally close solutions. This time these are solutions of the equation of Lagrangian trajectories. The solutions provide trajectories of fluid particles and form a three-dimensional dynamical system. This system is characterized by a positive Lyapunov exponent $\lambda_1$ that describes exponential growth of distance $r$ between two trajectories below the viscous scale \cite{frisch}, the phenomenon often referred to as the Lagrangian chaos, see e.g. \cite{fb,gaw,mj,reviewt,fk}.

%It was found in \cite{jeremy} that $Re-$dependence of $\lambda_1$, and the dispersion of the finite-time Lyapunov exponent, are described well by the laws predicted in \cite{cr} for $\lambda^v$ and the dispersion of the finite time Lyapunov exponent of turbulence. The authors have not commented on the reasons of successfully applying the theory for Lyapunov exponents of turbulence to those of fluid particles. This work implies that $\lambda_1\sim \langle t_{\nu}^{-1}\rangle$ provides a qualitatively valid description of the $Re-$dependence of the Lyapunov exponent. Indeed, we demonstrate that the calculation of \cite{cr}, originally intended for turbulence, in fact applies to the fluid particles. This is the reason why using the prediction of \cite{cr} for another quantity, \cite{jeremy} could explain their observations.
%PJK: linebreaks for legibility
It was found in \cite{jeremy} that $Re-$dependence of $\lambda_1$, and
the dispersion of the finite-time Lyapunov exponent, are described
well by the laws predicted in \cite{cr} for $\lambda^v$ and the
dispersion of the finite time Lyapunov exponent of turbulence. The
authors have not commented on the reasons of successfully applying the
theory for Lyapunov exponents of turbulence to those of fluid
particles. This work implies that $\lambda_1\sim \langle
t_{\nu}^{-1}\rangle$ provides a qualitatively valid description of the
$Re-$dependence of the Lyapunov exponent. Indeed, we demonstrate that
the calculation of \cite{cr}, originally intended for turbulence, in
%fact applies to the fluid particles. This is the reason why using the
%PJK: omit 1st 'the'
fact applies to fluid particles, cf. \cite{ho}. This is the reason why using the
prediction of \cite{cr} for another quantity, \cite{jeremy} could
explain their observations.

The growth of the distance between two close trajectories of fluid
particles is intermittent. There are two different sources of
%intermittency involved. One type of intermittency is well-known, see
%PJK: One => The first
intermittency involved.

The first type of intermittency is
well-known, see
%PJK
e.g. the detailed discussion in \cite{fk}. It does not have anything to do with the intermittency of turbulence. This intermittency originates in randomness of the velocity field and exists even for separation in Gaussian flows that are completely
uncorrelated in time, the so-called Kraichnan model \cite{reviewt}. One can understand it in the following way: At large times, much larger than the correlation time of the flow, the most probable value of the finite-time average stretching rate on the trajectory is $\lambda_1$. (See Eqs.~(\ref{log})-(\ref{sv}) below.) Thus $r\sim \exp(\lambda_1 t)$ holds with probability close to one. However, there are also rare trajectories for which the finite-time average stretching rate $\lambda$ on the trajectory is larger than  $\lambda_1$. For these trajectories the local velocity gradient is systematically larger than $\lambda_1$. This persistence of large gradients occurs not because the separating pair of particles entered a region of long living gradients of the flow. Rather, the gradients in question, that are qualitatively randomly renewed each correlation time, attain atypical value $\simeq \lambda$ after each renewal. The probability of these randomly persistent configurations of the flow decays exponentially in time. Despite the small probability of these rare events, they dominate moments of the inter-particle distance, because they are associated with separation $r\sim \exp(\lambda t)$, which is exponentially larger than $\exp(\lambda_1 t)$.
%This occurs when the
%velocity gradient on the trajectory of one of the separating particles
%deviates systematically from the typical value. This happens not
%because the trajectory is inside a region of persistent velocity
%gradients. In fact, it The probability of having systematic deviations decays
%in time exponentially becoming very small quickly. However these
%PJK: comma
%in time exponentially becoming very small quickly. However, these
%events are still non-negligible since for them the separation rate is
%larger than typical.
%As a result these rare events %produce
%%exponentially larger distance between the separating particles and
%%determine the moments of the distance \cite{fk}. This type of
%%intermittency is intermittency of separation that does not have
%%PJK: Can we just say 'This is intermittencey of separation...'
%determine the moments of the distance \cite{fk}.

%This
%is intermittency of separation that does not have
%%anything to do with the intermittency of turbulence and the Reynolds
%%PJK: and => or
%anything to do with the intermittency of turbulence \red{or} the Reynolds
%%PJK: I think that more needs to be said about the difference between
%%PJK: two types of intermittency for the average (or below) reader
%%PJK: (e.g. myself). More specifically, it is not very obvious what
%%PJK: exactly is the difference between the two types of intermittency.
%%PJK: Itzhak, could you perhaps add a sentence or two to expand this?
%number.

The other contribution to the intermittency of separation of particle trajectories, which is in fact the dominant contribution at large $Re$, is due to intermittency of turbulence. The probability of having atypically large persistent velocity gradients, where the finite-time average stretching rate exceeds $\lambda_1$ by a power of $Re$, is significant in turbulence, as opposed to Gaussian randomness. This results in magnification of growth rates by powers of $Re$, a phenomenon we call ``hyper-intermittency".

The intermittent separation of the trajectories cannot be described by $\lambda_1$ only and demands the introduction of the generalized Lyapunov exponent $\gamma(k)$. This provides the growth exponent of the $k-$th moment of the distance between the trajectories, see e.g.  \cite{mj} for numerical studies and \cite{fk} for the theory. We demonstrate that $\gamma(k)/k$ is qualitatively similar to $\gamma^v(k)$ where factor of $k$ is due to insignificant difference in the definitions. This similarity holds since both sets of the exponents are determined by similar processes of local stretching of fluid elements.

The bridge between the two sets of the exponents is provided by the growth rate of small fluctuations of magnetic field in the turbulent flow of a conducting fluid. The field reacts on the transporting turbulent flow via the Lorentz force. We consider only early stages of the magnetic field amplification by the turbulent flow from its infinitesimal seed values where the Lorentz force is negligible. The flow is then prescribed and unaffected by the dynamics of the magnetic field. It obeys the same Navier-Stokes equations as without the magnetic field which is the so-called kinematic dynamo regime \cite{ll8,xl}.

The field's growth in ideal conducting flow, defined by setting resistivity to zero, is fully described by $\gamma(k)$ since the magnetic field lines are ``frozen" in the fluid \cite{ll8}. Thus spatial moment of magnetic field of order $k$ grows in time exponentially with exponent $\gamma(k)$. The long-time asymptotic growth rate at almost every fixed point in space is uniform and given by the Lyapunov exponent $\lambda_1$ that equals $\gamma'(0)$. The dimensionless growth exponent $\lambda_1\tau_{\nu}$, where $\tau_{\nu}$ is the Kolmogorov time \cite{frisch}, decays with $Re$, see above. However this does not mean that similar dependence holds also for the energy whose dimensionless growth exponent  is given by $\gamma(2)\tau_{\nu}$. We demonstrate that due to intermittency $\gamma(2)\tau_{\nu}$ grows with $Re$ with appreciable scaling exponent, quite similarly to the growth of $\lambda^v \tau_{\nu}=\gamma^v(2)\tau_{\nu}$ observed in \cite{mohan}. The difference holds because the energy integral at time $t$ is determined by rare spatial regions where the growth exponent of the energy is larger than $\lambda_1$. The volume of these regions shrinks exponentially fast and disappears completely at $t\to\infty$ in accord with the uniform asymptotic growth at exponent $\lambda_1$. Still at any finite $t$ this volume is finite and it determines the energy integral, cf. \cite{fk}.

The hyper-intermittency of separation of trajectories has direct implications for the intermittent growth of the magnetic field. The energy grows due to regions whose volume fraction is exponentially small in time. This is true already in model Gaussian velocity fields, see e.g. \cite{falk,xl}. However for those fields the rate of growth in these regions is not larger than $\lambda_1$ by a power of a large parameter as $Re$ above. It is merely larger than $\lambda_1$ resulting in the inequality $\gamma(2)>2\lambda_1$ where $\gamma(2)/(2\lambda_1)$ is of order one (strictly speaking convexity allows for $\gamma(2)=2\lambda_1$ however this degenerate case $\gamma(k)=k\lambda_1$ does not seem relevant for random flows). In contrast the growth exponent in the intermittent turbulent flow $\gamma(2)/(2\lambda_1)$ is given by a power of $Re$ and is a larger parameter in the high-$Re$ flows. Thus due to hyper-intermittency the growth is exponentially sensitive both to time and $Re$. The observation of hyper-intermittent growth at large $Re$ poses a formidable challenge due to extreme rarity of the statistically relevant regions.

%Finally, before turning to quantitative studies, we describe how the passage from $\gamma(k)$ to $\gamma^v(k)$ can be realized. The data which is available presently and the theory below indicate that $\gamma(k)=c_kRe^{\beta_k}$ where $c_k$ are either independent of $Re$ or depend on it slowly and $\beta_k$ is a non-trivial function of $k$. We consider growth exponents of the magnetic field's moments $\gamma^B(k)$ as functions of magnetic diffusivity $\eta$ where $\gamma^B(k, \eta=0)=\gamma(k)=c_kRe^{\beta_k}$. The $Re-$dependence of $\gamma^B(k, \eta)$ at small but finite $\eta$ is similar to that of $\gamma(k)$ as can be seen from the solution of \cite{falk} at $\eta$ much smaller than the kinematic viscosity $\nu$. Thus, though, $\gamma^B(k, \eta)$ has a jump at $\eta=0$ this jump does not change the $Re-$dependence of the asymptotic growth rate $d\gamma^B/dk(k=0)$ and does not seem to change qualitatively the $Re-$dependence of other $\gamma(k)$. Furthermore it does not seem possible that the continuation from small finite $\eta/\nu$ to $\eta/\nu=1$ will change the $Re-$dependence of $\gamma^B(k)$ qualitatively. Hence $\gamma^B(k, \eta=\nu)$ depends on $Re$ as a power-law with $k-$dependent exponent. However the only difference of the equations of $\bm B$ at $\eta=\nu$ and $\delta \bm v$ is a sign of the stretching term and the presence of pressure. Both would not change the qualitative $Re-$dependence proposing that $\gamma^v(k)=b_kRe^{\delta(k)}$ where $Re-$dependence of $b_k$ is slow.
%PJK: Linebreaks for legibility
Finally, before turning to quantitative studies, we describe how the
passage from $\gamma(k)$ to $\gamma^v(k)$ can be realized. The data
which is available presently and the theory below indicate that
%$\gamma(k)=c_kRe^{\beta_k}$ where $c_k$ are either independent of $Re$
%or depend on it slowly and $\beta_k$ is a non-trivial function of
%PJK: changed order sligthly
$\gamma(k)=c_kRe^{\beta_k}$, where $c_k$ are either independent
  or weakly dependent on $Re$, and $\beta_k$ is a non-trivial
function of
%PJK
$k$. We consider growth exponents of the magnetic field's moments
$\gamma^B(k)$ as functions of magnetic diffusivity $\eta$ where
$\gamma^B(k, \eta=0)=\gamma(k)=c_kRe^{\beta_k}$. The $Re-$dependence
of $\gamma^B(k, \eta)$ at small but finite $\eta$ is similar to that
of $\gamma(k)$ as can be seen from the solution of \cite{falk} at
$\eta$ much smaller than the kinematic viscosity $\nu$. Thus, though,
$\gamma^B(k, \eta)$ has a jump at $\eta=0$ this jump does not change
the $Re-$dependence of the asymptotic growth rate $d\gamma^B/dk(k=0)$
and does not seem to change qualitatively the $Re-$dependence of other
$\gamma(k)$. Furthermore it does not seem possible that the
continuation from small finite $\eta/\nu$ to $\eta/\nu=1$ will change
the $Re-$dependence of $\gamma^B(k)$ qualitatively. Hence $\gamma^B(k,
\eta=\nu)$ depends on $Re$ as a power-law with $k-$dependent
exponent. However the only difference of the equations of $\bm B$ at
$\eta=\nu$ and $\delta \bm v$ is a sign of the stretching term and the
presence of pressure. Both would not change the qualitative
$Re-$dependence proposing that $\gamma^v(k)=b_kRe^{\delta(k)}$ where
$Re-$dependence of $b_k$ is slow.

\section{$Re-$dependence of generalized Lyapunov exponent of fluid particles}

We assume everywhere in this work that the flow $\bm v(t, \bm x)$ evolves according to the Navier-Stokes (NS) equations
\begin{eqnarray}&&\!\!\!\!\!\!\!\!\!\!\!\!\!\!
\partial_t\bm v+(\bm v\cdot\nabla)\bm v=-\nabla p+\nu\nabla^2\bm v+\bm f,\ \ \nabla\cdot\bm v=0, \label{ns}
\end{eqnarray}
where $p$ is the pressure divided by the density, $\nu$ is the kinematic viscosity and the forcing $\bm f$ ensures stationarity. The flow is characterized by the characteristic value $V$ of velocity at the integral scale $L$ and the Reynolds number $Re\equiv VL/\nu$. The flow, which is assumed to be turbulent, $Re\gg 1$, generates a three-dimensional dynamical system whose trajectories $\bm q(t, \bm x)$ obey
\begin{eqnarray}&&\!\!\!\!\!\!\!\!\!\!\!\!\!\!
\partial_t \bm q(t, \bm x)=\bm v(\bm q(t, \bm x), t),\ \ \bm q(t=0, \bm x)=\bm x. \label{ts}
\end{eqnarray}
Here $\bm q(t, \bm x)$ are Lagrangian trajectories of the fluid that are labeled by their position at $t=0$. We consider homogeneous turbulence though many considerations below hold for inhomogeneous turbulence as well. We assume that the flow $\bm v$ is smooth below the viscous scale $l_{\nu}$. Then the above system is a smooth dynamical system which can be characterized by the Lyapunov exponent. The distance $\bm r(t, \bm x)\equiv \bm q(t, \bm x+\bm r_0)- \bm q(t, \bm x)$ between two Lagragian trajectories initially separated by $r_0\ll l_{\nu}$ obeys
\begin{eqnarray}&&\!\!\!\!\!\!\!\!\!\!\!\!\!\!
\partial_t \bm r(t, \bm x)=\bm v(\bm q(t, \bm x)+\bm r, t)-\bm v(\bm q(t, \bm x), t)\approx (\bm r\cdot \nabla)\bm v, \label{s}
\end{eqnarray}
where $\nabla\bm v$ is evaluated at $\bm q(t, \bm x)$ and we consider not too large times such that $r(t)\ll l_{\nu}$. We concentrate on the distance $r(t)$ by introducing $\bm r=r{\hat n}$ where $|{\hat n}|=1$. We find from Eq.~(\ref{s}),
\begin{eqnarray}&&\!\!\!\!\!\!\!\!\!\!\!\!\!\!
\frac{d\ln r}{dt}={\hat n}\nabla \bm v {\hat n},\ \ \frac{d{\hat n}}{dt}=({\hat n}\cdot\nabla)\bm v-{\hat n}({\hat n}\nabla \bm v {\hat n}).
\end{eqnarray}
This gives
\begin{eqnarray}&&\!\!\!\!\!\!\!\!\!\!\!\!\!\!
\frac{1}{t}\ln \left(\frac{r(t, \bm x)}{r_0}\right)\!=\!\int_0^t {\hat n}(t')\nabla \bm v (t'){\hat n}(t')\frac{dt'}{t}, \label{log}
\end{eqnarray}
where we omitted the $\bm x-$dependence in the RHS. The Lyapunov exponent then describes the limit
\begin{eqnarray}&&\!\!\!\!\!\!\!\!\!\!\!\!\!\!
\lim_{t\to\infty}\frac{\ln (r(t, \bm x)/r_0)}{t}\!=\!\lim_{t\to\infty}\int_0^t {\hat n}\nabla \bm v {\hat n}\frac{dt'}{t} \!\equiv\! \lambda_1(\bm x). \label{sv}
\end{eqnarray}
In the case of time-independent flows the time-average above is described by the Oseledets theorem \cite{oseledets}, sometimes called the multiplicative ergodic theorem. Its generalization to random flows, considered here,
states that the limit exists and is independent of $\bm x$ and the realization of the velocity \cite{gaw}. We designate the corresponding constant by $\lambda_1$. The independence holds with probability one that is except for set of $\bm x$ with zero total volume (below ``for almost all $\bm x$" or ``almost everywhere" abbreviated as a. e.) and for velocity fields with zero total probability measure. Here the probability is defined by the so-called natural measure which corresponds to averaging over both $\bm x$ and the realization of the flow \cite{gaw}. Qualitatively, similarly to the usual ergodic theorem, the multiplicative ergodic theorem means that ${\hat n}\nabla \bm v {\hat n}$ can be considered as a stationary random process with a finite correlation time so that the law of large numbers holds.

We observe that since $\lambda_1(\bm x)=\lambda_1$ with probability one then $\lambda_1$ is a self-averaging quantity. It can be obtained by averaging Eq.~(\ref{sv}) over $\bm x$ and velocity statistics.  We start from space averaging
\begin{eqnarray}&&\!\!\!\!\!\!\!\!\!\!\!\!\!\!
\lambda_1\!=\!\lim_{t\to\infty} \frac{1}{t} \int d\bm x\int_0^t dt' {\hat n}\nabla \bm v {\hat n},
\end{eqnarray}
where the volume was set to one by choice of units of length. We observe that ${\hat n}(t)$ relaxes exponentially for almost all $\bm x$ to a unique direction that is independent of the initial condition ${\hat n}(0)$. The characteristic relaxation time is few $\tau_{\nu}$, see e.g. \cite{fp} for detailed consideration. This unique direction, called below the major stretching direction, defines a field ${\hat n}(t, \bm x)$. This field can be defined at any $t$ by considering long evolution of ${\hat n}$ on Lagrangian trajectories that arrive at $\bm x$ at time $t$, or, more directly, by studying backward in time evolution. Since we consider infinite-time limit while ${\hat n}$ settles on the major stretching field ${\hat n}(t,\bm x)$ after few Kolmogorov times $\tau _{\nu}$ then
\begin{eqnarray}&&\!\!\!\!\!\!\!\!\!\!\!\!\!\!
\lambda_1\!=\!\lim_{t\to\infty} \frac{1}{t} \int d\bm x\int_0^t dt' \left[{\hat n}\nabla \bm v {\hat n}\right](t', \bm q(t', \bm x))
\nonumber\\&&\!\!\!\!\!\!\!\!\!\!\!\!\!\!
=\!\lim_{t\to\infty} \frac{1}{t} \int d\bm x\int_0^t dt' {\hat n}(t', \bm x)\nabla \bm v(t', \bm x) {\hat n}(t', \bm x), \label{las}
\end{eqnarray}
where we changed integration variable from $\bm x$ to $\bm q(t', \bm x)$ keeping the notation for the integration variable with no ambiguity. This quantity by the property of $\lambda_1(\bm x)$ described above must be  independent of the realization of the flow if the realization is typical. Indeed, the space-time average in Eq.~(\ref{las}) is of the type which is known to be independent of the flow. This is often used in turbulence studies for numerical calculations of averages over ensemble of velocities by using space-time averaging instead, see \cite{lee} and references therein. Thus the above representation allows to obtain $\lambda_1$ numerically by using only one realization of the flow. One needs to average ${\hat n}\nabla \bm v {\hat n}$ at random points in space and time where here and below ${\hat n}$ is understood as the local major stretching direction.

Since the last term in Eq.~(\ref{las}) is independent of the velocity realization, averaging this equation over the velocity ensemble we find the representation of $\lambda_1$ as velocity average, $\lambda_1=\left\langle {\hat n}\nabla \bm v {\hat n}\right\rangle$. Here the remaining space-time averaging can be omitted since the velocity average is a constant in space and time stationarity and spatial homogeneity of turbulence. Here and below the angular brackets without subscript stand for the velocity ensemble average.

We demonstrate how the representation of $\lambda_1$ as velocity average of ${\hat n}\nabla \bm v {\hat n}$, with ${\hat n}$ the major stretching direction, could be also obtained by first averaging over the velocity and then over space. We have after averaging Eq.~(\ref{sv}) over the velocity
%independent of realization of velocity and hence providing the result of averaging over the ensemble of velocities. We see that  The above implies
%Thus $\lambda_1=\left\langle {\hat n}\nabla \bm v {\hat n}\right\rangle$ We obtain the same result if we
\begin{eqnarray}&&\!\!\!\!\!\!\!\!\!\!\!\!\!\!
\lambda_1\!=\!\lim_{t\to\infty}\int_0^t \left\langle {\hat n}\nabla \bm v {\hat n}\right\rangle\frac{dt'}{t}=\lim_{t\to\infty} \left\langle  {\hat n}(t)\nabla \bm v (t){\hat n}(t)\right\rangle,
\end{eqnarray}
where we use that at large times the process ${\hat n}\nabla \bm v {\hat n}$ is stationary due to relaxation of ${\hat n}$ to the major stretching direction and the average is constant. Since the ensemble average above is independent of $\bm x$ for homogeneous turbulence then further averaging over $\bm x$, that must be done in principle \cite{gaw}, is redundant. Thus we recover that spatio-temporal average in Eq.~(\ref{las}) must be equal to velocity ensemble average $\left\langle  {\hat n}\nabla \bm v {\hat n}\right\rangle$ as necessary for the self-consistency of the assumption made after Eq.~(\ref{las}).
%Here and below the angular brackets are used interchangeably for space and velocity averaging where it is clear from the context which one is used. We went to some length here since there seems to be no detailed exposition of these issues in the literature.

%The multiplicative ergodic theorem allows to find $\lambda_1$ as average over the statistics of velocity $\lambda_1=\left\langle {\hat n}\nabla \bm v {\hat n}\right\rangle$ with some refinement explained below (We observe that since \st{Here we use} {\color{red}{Recalling}} that the natural measure is the one on space and velocities {\color{em}{not clear statement}} \cite{gaw} \st{and} {\color{red}{we}} consider homogeneous turbulence (generalization to inhomogeneous case can be considered).

The main conclusion from the above that is needed below is that $\lambda_1$ equals the average over the velocity ensemble $\left\langle  {\hat n}\nabla \bm v {\hat n}\right\rangle$ where ${\hat n}$ is the major stretching direction. For turbulence $\lambda_1$ is positive, the fact that defines Lagrangian chaos of motion of fluid particles.

\subsection{Dependence of the Lyapunov exponent on $Re$} \label{lyapunov}

We consider the Reynolds number dependence of the dimensionless Lyapunov exponent $\lambda_1\tau_{\nu}$ where $\tau_{\nu}\equiv \sqrt{\nu/\epsilon}$, defined by the energy dissipation rate per unit volume $\epsilon$, is the Kolmogorov time \cite{frisch}. Using incompressibility and spatial homogeneity we write
\begin{eqnarray}&&\!\!\!\!\!\!
\frac{\epsilon}{\nu}\!=\left\langle (\nabla_iu_k)(\nabla_iu_k)\right\rangle\!=\!\left\langle (\nabla_iu_k\!+\!\nabla_ku_i)\nabla_iu_k\right\rangle\!=\!2\left\langle tr s^2\right\rangle,\nonumber
\end{eqnarray}
where $s_{ik}\equiv (\nabla_iv_k+\nabla_kv_i)/2$ is the rate-of-strain matrix. It should be emphasised that $\tau _{\nu}$ introduced here is a parameter that characterises the entire flow and known as the Kolmogorov time, and not the fluctuating local viscous time $t_{\nu}$ considered in the Introduction.

{\bf Inequalities on $\lambda_1$}--- We introduce the ordered eigenvalues $s_i$ of the symmetric rate-of-strain matrix, $s_1\geq s_2\geq s_3$. The incompressibility condition implies that $s_1+s_2+s_3=0$ and therefore that $s_1\geq 0$ and $s_3\leq 0$. The first inequality may be derived from the relationship $\lambda _1=\left\langle {\hat n}s{\hat n}\right\rangle$, which yields
\begin{eqnarray}&&\!\!\!\!\!\!\!\!\!\!\!\!\!\!
\lambda_1^2\leq \left\langle \left({\hat n}s{\hat n}\right)^2\right\rangle\leq \left\langle max_{i=1}^3\  s_i^2\right\rangle.%\nonumber
%\\&&\!\!\!\!\!\!\!\!\!\!\!\!\!\!
%=\sum_{i=1}^3\left\langle \lambda_i^2\right\rangle=2\left\langle \left(\lambda_1^2+\lambda_3^2+\lambda_1\lambda_3\right)\right\rangle\geq 2\left\langle \left(\lambda_1^2+\lambda_3^2\right)\right\rangle\geq 6\left\langle \lambda_1^2\right\rangle.\nonumber
\end{eqnarray}
We observe that $|s_2|\leq s_1$ and $|s_2|\leq |s_3|$. Thus $max_i\  s_i^2$ is either $s_1^2$ or $s_3^2$. If $max_i\  s_i^2=s_1^2$ then $s_2<0$ and $s_2^2+s_3^2\geq s_1^2/2$. Here we rely on the result that the minimum of $s_2^2+s_3^2$, subject to the incompressibility condition $|s_2|+|s_3|=s_1$ is attained at $|s_2|=|s_3|=s_1/2$. This results in $tr s^2=\sum_{i=1}^3s_i^2 \geq 3 s_1^2/2=3\ max_i\  s_i^2/2$. Performing similar consideration in the case where $max_i\  s_i^2=s_3^2$ by sign reversal of $s_i$ we conclude that
\begin{eqnarray}&&\!\!\!\!\!\!\!\!\!\!\!\!\!\!
\frac{3\   max_{i=1}^3\  s_i^2}{2}\leq tr s^2. \label{inq}
\end{eqnarray}
We find by comparing with $\epsilon/\nu$ the inequality
\begin{eqnarray}&&\!\!\!\!\!\!\!\!\!\!\!\!\!\!
\lambda_1^2\leq \left\langle max_i\  s_i^2\right\rangle\leq \frac{2 \left\langle tr s^2\right\rangle}{3}=\frac{\epsilon}{3\nu},
\end{eqnarray}
or
\begin{eqnarray}&&\!\!\!\!\!\!
\lambda_1\tau_{\nu}\leq \frac{1}{\sqrt{3}}=0.577, \label{ourin}
\end{eqnarray}
cf. \cite{ym}. For $Re$ that are accessible by today's simulations it is found that $\lambda_1\tau_{\nu}$ is about $0.14$, see \cite{jeremy} and cf. \cite{mj}. The value of the above inequality is that it demonstrates that $\lambda_1\tau_{\nu}$ is bounded from above and cannot grow with the Reynolds number indefinitely, as by a power-law or otherwise, see below.

A second and stronger inequality may be derived by writing $\lambda_1\leq \left\langle \left|{\hat n}s{\hat n}\right|\right\rangle\leq \left\langle max_{i=1}^3 \left|s_i\right|\right\rangle$. Since $max_{i=1}^3 \left|s_i\right|=\sqrt{max_i\  s_i^2}$,  we find from Eq.~(\ref{inq}) that
\begin{eqnarray}&&\!\!\!\!\!\!
\lambda_1\tau_{\nu}\leq \frac{\left\langle \sqrt{{\tilde \epsilon}}\right\rangle}{\sqrt{3}},\ \ {\tilde \epsilon}\equiv \frac{tr s^2}{\left\langle tr s^2\right\rangle}, \label{in}
\end{eqnarray}
where we introduced the normalized dissipation ${\tilde \epsilon}$. This inequality allows to derive a power-law decay of $\lambda_1\tau$ with $Re$ if we can make the usual assumption on the power-law dependence of the moments of dissipation on $Re$ caused by the intermittency,
\begin{eqnarray}&&\!\!\!\!\!\!
\left\langle {\tilde \epsilon}^k\right\rangle\sim Re^{\sigma(k)}, \label{phso}
\end{eqnarray}
see e. g. \cite{frisch,sreeni,lee}. It is readily seen from H\"{o}lder's inequality that $\sigma(k)$ is convex.
%Then the constraints $\sigma(0)=\sigma(1)=0$, implied by the definition in Eq.~(\ref{phso}) mean that $\sigma(k)$ is non-positive for $0<k<1$ and non-negative otherwise \cite{lee}. Thus $\sigma(1/2)\leq 0$ and, assuming that there is no degeneracy and the inequality is strict, $\sigma(1/2)<0$, we find from Eq.~(\ref{in}) that $\lambda_1\tau_{\nu}$ is bounded from above by a power-law function of $Re$ with negative exponent.
%PJK: linebreaks for legibility
Then the constraints $\sigma(0)=\sigma(1)=0$, implied by the
%definition in Eq.~(\ref{phso}) mean that $\sigma(k)$ is non-positive
%PJK: non-positive => negative
definition in Eq.~(\ref{phso}) mean that $\sigma(k)$ is non-positive
%for $0<k<1$ and non-negative otherwise \cite{lee}. Thus
%PJK: non-negative => positive
for $0<k<1$ and non-negative otherwise \cite{lee}. Thus
$\sigma(1/2)\leq 0$ and, assuming that there is no degeneracy and the
inequality is strict, $\sigma(1/2)<0$, we find from Eq.~(\ref{in})
that $\lambda_1\tau_{\nu}$ is bounded from above by a power-law
function of $Re$ with negative exponent.

It is highly plausible that the bound holds as an order of magnitude equality so that
\begin{eqnarray}&&\!\!\!\!\!\!\!\!\!
\lambda_1=\left\langle {\hat n}\nabla \bm v{\hat n}\right\rangle\sim \left\langle \sqrt{(\nabla\bm v)^2}\right\rangle,\ \ \lambda_1\tau_{\nu}=c_0Re^{\sigma(1/2)},
\end{eqnarray}
where $c_0$ is either a constant or a function that depends on $Re$ slower than a power-law. For simple approximate calculation of $\sigma(1/2)$ see \cite{lee}.

{\bf Prediction of multifractal model}---In the frame of phenomenology of turbulence \cite{frisch} velocity gradients are estimated as $t_{\nu}^{-1}$ where $t_{\nu}$ is the fluctuating viscous time introduced before. We find then $\lambda_1=\left\langle {\hat n}\nabla \bm v{\hat n}\right\rangle\sim\langle t_{\nu}^{-1}\rangle$. The last average was calculated using the multifractal model \cite{frisch} in \cite{cr} yielding
\begin{eqnarray}&&\!\!\!\!\!\!
\lambda_1\tau_{\nu}\sim Re^{-\delta}, \label{prediction}
\end{eqnarray}
with $\delta\simeq 0.041$ (below we often refer to the equation above where $\delta$ is considered as a phenomenological constant, not necessarily equal to $0.041$). This prediction, understood as the prediction for $\lambda_1$ and not $\lambda_1^v$, was confirmed qualitatively in \cite{jeremy}. The work observed slow robust decay of $\lambda_1\tau_{\nu}$ with the Taylor microscale Reynolds number $Re_{\lambda}$ taking values of $65$, $105$ and $185$. Therefore we seem to have all indications that $\lambda_1\tau_{\nu}$ decays with $Re$ as a power-law, albeit with a small exponent. This smallness is significant because $Re$ could get very large in applications. If $\delta=0.041$ of \cite{cr} is accepted in Eq.~(\ref{prediction}), then $\lambda_1\tau_{\nu}$ is of order $10^{-1}$ at all practically relevant $Re$, including $Re$ as high as $10^{15}$ appearing in astrophysical applications.

%{\bf Resolution of the apparent contradictions}---
We explained in the Introduction that \cite{cr} assumed $\lambda^v\sim \langle t_{\nu}^{-1}\rangle$. Correspondingly they predicted $\lambda^v\tau_{\nu}\sim Re^{-\delta}$ which as we discussed contradicts the observations. Our consideration above shows that by itself the calculation of \cite{cr} is useful.
%The authors of \cite{cr} assumed that the above consideration holds for $\lambda_1^v$ and proposed that $\lambda_1^v\tau_{\nu}\sim Re^{-\delta}$. This was found to be in contradiction with numerical observations of \cite{mohan} who found that $\lambda_1^v\tau_{\nu}$ grows with the Reynolds number according to a power-law $Re^{\beta}$ with $\beta$ between $1/8$ and $1/6$. {\color{red}{The authors of \cite{mohan} have raised the possibility that this is due to the small values of the Reynolds number and for higher values $\lambda _1$ indeed saturates}} \st{This could tell that their observation is a low Reynolds number phenomenon, as was conjectured by the authors as a possible reason for their observation} \st{(the observed value of $\lambda_1^v\tau_{\nu}=0.16$ agrees with}\sout{ Eq.~(\ref{ourin})}. We demonstrate below however that $\lambda_1$ and $\lambda_1^v$ are different quantities. \st{It is wrong to assume} {\color{red}{We show}} that $\lambda_1^v$ is {\color{red}{not related to}} the average of the inverse viscous time-scale $t_{\nu}$, rather
%$\lambda_1^v\sim t^{-1} \ln \left\langle \exp(2t/t_{\nu})\right\rangle$ with $t\to\infty$ holds qualitatively.

{\bf Remarks}---The above derivations assume
%We presented above two derivations of Eq.~(\ref{prediction}). One derivation, which is reinterpreted derivation of \cite{cr}, allows to evaluate $\delta$. The other derivation relies on Eqs.~(\ref{in})-(\ref{phso}) and the assumption of non-degeneracy of $\sigma(k)$, not allowing to fix $\delta$ {\color{em}{I do not see a principle difference between the possibility to determine $\delta$ in both derivations. We discussed in on Thursday but do not remember what we decided}}. Both of these derivations
multifractality, intermittency and breakdown of the Kolmogorov $1941$ theory's posit of self-similarity of turbulence. These assumptions are confirmed by absolute majority of the measurements that exist today, see e.g. the references in \cite{frisch,sreeni} for Eq.~(\ref{phso}) with $\sigma(k)\neq 0$. However there exists an alternative view that Eq.~(\ref{phso}) is only a finite $Re$ effect and at larger $Re$ at least some moments of ${\tilde \epsilon}$ become constants, independent of $Re$, much in agreement with the Kolmogorov theory \cite{antonia}. This view seemingly would imply that $\lambda_1\tau_{\nu}\sim Re^{-\delta}$ is only an intermediate law applying at finite $Re$ and $\lim_{Re\to\infty} \lambda_1\tau_{\nu}$ is finite. Further the derivation of $\lambda_1\tau_{\nu}\sim Re^{-\delta}$ by \cite{cr} could also be criticized since it uses the phenomenology of turbulence that was criticized recently in \cite{bud} who propose that the phenomenology does not hold at least until the Taylor-microscale Reynolds number of order $10^4$. However \cite{bud} would seemingly confirm Eq.~(\ref{phso}) so that our derivation would work. We conclude that it is highly plausible that $\lambda_1\tau_{\nu}\sim Re^{-\delta}$ with a small, non-zero $\delta$ is valid.

Qualitatively the decay of $\lambda_1\tau_{\nu}$ with $Re$ holds because
%quiescent regions of turbulence, where chaos is depleted, become longer in time and larger in space due to intermittency \cite{frisch}. The increase of quiescent regions is accompanied by increase in the amplitude of the bursts however it is the former that determine $\lambda_1$.
%PJK: linebreaks for legibility
quiescent regions of turbulence, where chaos is depleted, become
longer in time and larger in space due to intermittency
\cite{frisch}. The increase of quiescent regions is accompanied by
%increase in the amplitude of the bursts however it is the former that
%PJK: Break sentence
increase in the amplitude of the bursts. However, it is the
former that determine $\lambda_1$.

{\bf Other Lyapunov exponents}---The behavior of $\lambda_1$ has implications for other Lyapunov exponents $\lambda_i$. The Lyapunov exponents of the fluid particles in the dissipation range of turbulence, where the ordering $\lambda_i\geq \lambda_{i+1}$ is assumed, are defined so that $\lambda_1+\lambda_2$ is the logarithmic growth rate of infinitesimal area elements and $\lambda_3=-\lambda_1-\lambda_2$ (more generally $\sum_{i=1}^3\lambda_i$ is the logarithmic rate of growth of infinitesimal volumes which vanishes for incompressible flows, see \cite{review,fm} for definitions). We observe that $\lambda_3=-\lambda_1-\lambda_2$ and $\lambda_i\geq\lambda_{i+1}$ give the inequality $-2\lambda_1\leq \lambda_3\leq -\lambda_1/2$ that implies that $\lambda_3$ (and thus also $\lambda_2$) must obey similar asymptotic dependence on the Reynolds number (presently available simulations reveal that $\lambda_2\simeq\lambda_1/4$ and $\lambda_3\simeq -5\lambda_1/4$, see e. g. \cite{mj}). This conclusion is also seen from the
%observation that dependence of $\lambda_3$ on the statistics of velocity gradients in very similar to that of $\lambda_1$, see Appendix of \cite{fb}.
%PJK: in => is
observation that dependence of $\lambda_3$ on the statistics of
velocity gradients is very similar to that of $\lambda_1$, see
Appendix of \cite{fb}.

\subsection{Generalized Lyapunov exponent}

We saw above that the dimensionless first Lyapunov exponent, that describes logarithmic growth rate of $r(t, \bm x)$, depends on $Re$ in observable yet not that strong way. Here we demonstrate that similar dependence for the growth exponent of other moments of $r(t, \bm x)$ can be strong. These moments are physically relevant e.g. the growth exponent of $\langle r^2(t)\rangle$ describes the growth of magnetic energy that will be described later and is also more similar to $\lambda^v$ than $\lambda_1$. Here we consider growth exponent $\gamma(k)$ of the $k-$th moment and describe its dependence on $Re$ (here $k$ can assume any real value not just integer values).

{\bf Definition and properties of generalized Lyapunov exponent}---We define the so-called generalized Lyapunov exponent $\gamma(k)$. This exponent describes the growth of moments of the distance between two infinitesimally close trajectories and can be introduced via
\begin{eqnarray}&&\!\!\!\!\!\!\!\!\!\!\!\!\!\!
\gamma(k)=\lim_{t\to\infty}\!\frac{\ln \left(\langle r^k(t)\rangle_s/r_0^k\right)}{t},\label{de}
\end{eqnarray}
where the limit $r_0\to 0$ is assumed to be taken before $t\to\infty$ so that $r(t)\ll l_{\nu}$ at all relevant $t$, cf. \cite{mj}. Equivalently $r(t)=r_0\exp\left(\int_0^t {\hat n}\nabla \bm v {\hat n}dt' \right)$ is used in the above equation, see Eq.~(\ref{log}). The subscript $s$ next to the angular brackets in Eq.~(\ref{de}) denotes spatial averaging over $\bm x$ so that
\begin{eqnarray}&&\!\!\!\!\!\!\!\!\!
\langle r^k(t)\rangle_s\equiv \int r^k(t, \bm x) d\bm x=r_0^k\left\langle \exp\left(k\int_0^t {\hat n}\nabla \bm v {\hat n}dt' \right)\right\rangle_s
\nonumber\\&&\!\!\!\!\!\!\!\!\! \sim r_0^k \exp(\gamma(k)t). \label{iwn}
\end{eqnarray}
The asymptotic equality holds at times much larger than the correlation time of the stationary process ${\hat n}\nabla \bm v {\hat n}$, attained after ${\hat n}(t)$ relaxes to the major stretching direction, see below.

It is by no means obvious that the limit in Eq.~(\ref{de}) gives a quantity that is independent of the realization of the flow. Indeed, the general formalism of random flows \cite{gaw} instructs us to average both over space and the ensemble of the velocities. For instance, the spatial average $\langle r^2(t)\rangle_s\equiv \int  r^2(t, \bm x) d\bm x$ at finite $t$ will vary from realization to realization of the flow. This is so because a given realization of the velocity field cannot produce all possible values of the RHS of Eq.~(\ref{log}) in finite time. However in the infinite time limit a given realization will almost surely scan through all possible values of
the RHS of Eq.~(\ref{log}). Thus we anticipate that  $\gamma(k)$ as defined in Eq.~(\ref{de}) should be independent of the realization of the flow. We shall now provide strong empirical arguments in favor of this realization-independence:

{\bf Cumulant series and realization independence of $\gamma(k)$}---A representation of $\gamma(k)$ as series in cumulants was introduced in \cite{fk}. Using the cumulant expansion theorem \cite{ma} we obtain:
\begin{eqnarray}&&\!\!\!\!\!\!\!\!\!\!\!\!\!\!
\gamma(k)\!=\!\lim_{t\to\infty}\!\frac{1}{t}\ln \left\langle \exp\left(k\int_0^t\!\! {\hat n}\nabla \bm v {\hat n}dt' \right)\right\rangle_s\!=\!k\lambda_1\!+\!\sum_{n=2}^{\infty}\frac{k^n}{n!} \nonumber\\&&\!\!\!\!\!\!\!\!\!\!\!\!\!\!
 \lim_{t\to\infty}\!\frac{1}{t} \int_0^t \!\!\left\langle{\hat n}\nabla\bm v{\hat n}(t_1){\hat n}\nabla\bm v{\hat n}(t_2)\ldots {\hat n}\nabla\bm v{\hat n}(t_{n})\right\rangle_s^c \prod_{i=1}^{n} dt_i,\label{cs}
\end{eqnarray}
where ${\hat n}$ is the major stretching direction and the superscript $c$ stands for cumulant. We remind the reader that the cumulant of the product of $n$ random variables, such as ${\hat n}\nabla\bm v{\hat n}(t_{i})$ above, is defined by taking out of the average all correlations of lower order. Thus for three random variables $x$, $y$ and $z$ we have $\langle xyz\rangle^c=\langle xyz\rangle-\langle x\rangle \langle yz\rangle-\langle y\rangle \langle xz\rangle-\langle z\rangle \langle xy\rangle+2 \langle x\rangle \langle y\rangle\langle z\rangle$. This implies that the cumulant in the integrand of the last line of the above equation is appreciable only when all $t_i$ are close within the correlation time of ${\hat n}\nabla\bm v{\hat n}(t)$, see details in \cite{ma}.

We see that $\gamma(k)$, the infinite-time limit of the cumulant generating function, is given by a series whose terms are space-time averages quite similar to the second average in Eq.~(\ref{las}). Following the same logic as we did after Eq.~(\ref{las}), these terms can be taken as realization-independent. These averages can be assumed \cite{fk} to be equal to the averages over the ensemble of velocities according to
\begin{eqnarray}&&\!\!\!\!\!\!\!\!\!\!\!\!\!\!
\gamma(k)=k\lambda_1+\frac{ k^2\mu}{2}+\sum_{n=3}^{\infty}\frac{k^n\Delta_n}{n!},
\label{cumulants} \\&&\!\!\!\!\!\!\!\!\!\!\!\!\!\!
\Delta_n\!\equiv \!\int_{-\infty}^{\infty}\!\! \!\left\langle{\hat n}\nabla\bm v{\hat n}(0){\hat n}\nabla\bm v{\hat n}(t_1)\ldots {\hat n}\nabla\bm v{\hat n}(t_{n-1})\right\rangle^c \prod_{i=1}^{n-1} dt_i.\nonumber
%\frac{k^3 }{3!}\int dt_1 dt_2  \times \langle{\hat n}\nabla\bm v{\hat n}(0){\hat n}\nabla\bm v{\hat n}(t_1) {\hat n}\nabla\bm v{\hat n}(t_2)\rangle_c
\end{eqnarray}
%In practice the velocity average is often found as space-time average, see the discussion of $\lambda_1$ above.
We introduced in Eq.~(\ref{cumulants}) the standard notation for the second order cumulant \cite{cr,jeremy}
\begin{eqnarray}&&\!\!\!\!\!\!\!\!\!\!\!\!\!\!
\mu=\int_{-\infty}^{\infty} \left\langle \left({\hat n}\nabla\bm v{\hat n}(0)-\lambda_1\right)\left({\hat n}\nabla\bm v{\hat n}(t)-\lambda_1\right)\right\rangle dt. \label{dam}
\end{eqnarray}
Thus we see that indeed $\gamma(k)$ is independent of velocity realization because it provides growth accumulated over infinite time. This is so because space averaging is always accompanied in Eq.~(\ref{cs}) by time-average, in contrast to $\langle r^k(t)\rangle_s$.
%
%As a result it is given by velocity-independent space-time average and not by a space average depending on velocity as $\langle r^k(t)\rangle_s$ considered above.

We shall demonstrate in Section \ref{mag} that $\gamma(k)$, for any fixed $k$, can be considered as a Lyapunov exponent of an infinite dimensional dynamical system consisting of the magnetic field in ideally conducting fluid. The $k-$dependence of the Lyapunov exponent arises because of different norms in functional space, in much the same way as $\gamma^v(p)$ which we discussed in the Introduction. We see that also in this interpretation $\gamma(k)$, as a Lyapunov exponent, is a realization-independent quantity.

To summarize, $\gamma(k)$ in Eq.~(\ref{de}) is well-defined. In particular, we mention its salient properties (see \cite{fk} and references therein): This function is convex, has two zeros at $k=-3$ and $k=0$, is negative for $-3<k<0$ and positive elsewhere. Note also $\gamma'(0)=\lambda_1>0$ and $\gamma'(-3)=\lambda_3<0$.

{\bf Dispersion $\mu$}---Multiplication of the series in Eq.~(\ref{cumulants}) by $\tau_{\nu}$ renders all terms dimensionless. In addition, they are non-trivial functions of $Re$. The most well-studied quantity, besides $\lambda_1\tau_{\nu}$ discussed above, is $\mu\tau_{\nu}$. In contrast to $\lambda_1\tau_{\nu}$, it is a growing function of $Re$. It was found in \cite{jeremy} that $\mu\tau_{\nu}$ depends on $Re$ significantly stronger than $\lambda_1\tau_{\nu}$. These authors observed that this result agrees with the calculation of the RHS of Eq.~(\ref{dam}) by \cite{cr} that was done by using the shell model of turbulence (it must be remarked though that \cite{cr} interpreted this quantity not as the dispersion $\mu$ but rather as analogous dispersion describing the growth of distance between solutions of the Navier-Stokes equations). The simulations of \cite{cr} gave $\mu\tau_{\nu}\sim Re^{\kappa}$ with
%PJK
$\kappa\sim 0.3$.
These authors elucidated the reason for this rather
strong growth of $\mu$ with $Re$. They argue that the equal-time correlation
function in the integrand of Eq.~(\ref{dam}) is not influenced by
intermittency. It is proportional to energy dissipation divided by the
viscosity and thus scales as $Re$ (up to the weak $Re-$dependence of
$\lambda_1\tau_{\nu}$, that enters the definition of $\mu$, that can
be disregarded). Therefore the dependence of $\mu\tau_{\nu}$ on $Re$
arises due to long correlation times of moderate gradients. Due to
intermittency the periods of calm turbulence become longer as $Re$
grows so that the correlation time behaves as a power of $Re$
leading to $\mu\tau_{\nu}\sim Re^{\kappa}$ law. Thus $\mu\tau_{\nu}$ grows as a power of $Re$ not because the local gradients
exceed the typical value, but rather because the typical gradients are
persistent over very long times.

It must be said that we are in a much worse position for estimating $\mu$ theoretically as compared to $\lambda_1$. As was discussed above we can write $\lambda_1\sim \langle t_{\nu}^{-1}\rangle$ where $\langle t_{\nu}^{-1}\rangle$ can be estimated using the multifractal model.  There is no multifractal or similar modeling that would work for $\mu$. The reason is that the multifractal model assumes validity of the phenomenology of turbulence \cite{frisch}. Within that phenomenology the lifetime
of a flow configuration with a given local stretching rate $|{\hat n}\nabla\bm v{\hat n}|$ is given by the inverse of that rate (as estimated from taking derivative of the Navier-Stokes equations giving $\partial_t\nabla \bm v=-(\nabla \bm v)^2+\ldots$). This would in fact predict that $\mu$ in Eq.~(\ref{dam}) behaves roughly as $|{\hat n}\nabla\bm v{\hat n}|$ and is similar to $\lambda_1$, which it is not. An adequate description requires a refinement of the phenomenology which also incorporates large fluctuations of lifetimes of configurations with a given $|{\hat n}\nabla\bm v{\hat n}|$.

{\bf Relevance of our results for numerical computation of cumulants}---For concreteness, let us demonstrate this relevance for the computation of $\mu$. It is preferable to use spatio-temporal averaging, which is more feasible practically than averaging over the velocity ensemble. We see from Eq.~(\ref{cs}) that after finding $\lambda_1$ we can obtain $\mu$ as the $t\to\infty$ limit of
\begin{eqnarray}&&\!\!\!\!\!\!\!\!\!\!\!\!\!\!
\mu\!=\!\!\int d\bm x
\int_0^t\!\! \left({\hat n}\nabla\bm v{\hat n}(t_1, \bm q(t_1, \bm x))\!-\!\lambda_1\right)
\nonumber\\&& \!\!\!\!\!\!\!\!\!\!\!\!\!\!\times
\left({\hat n}\nabla\bm v{\hat n}(t_2, \bm q(t_2, \bm x))\!-\!\lambda_1\right)\frac{ dt_1dt_2  }{t}.\label{space}
\end{eqnarray}
If we ignore intermittency the limit converges at rather small $t$. We observe that the value of ${\hat n}(0, \bm x)$ is determined by velocity gradients on the trajectory $\bm q(t, \bm x)$ (with $t<0$ in Eq.~(\ref{ts})) within the preceding time interval of order $\tau_{\nu}$ (that is $-\tau_{\nu}\lesssim t<0$). Here $\tau_{\nu}$ is the relaxation time of ${\hat n}(t)$ as we explained previously. This implies that the correlation length of ${\hat n}(0, \bm x)$ is the viscous scale $l_{\nu}$, since this is the scale over which
%$\nabla\bm v(t, \bm q(t, \bm x)$
%PJK: added missing bracket
$\nabla\bm v(t, \bm q(t, \bm x))$,
with $-\tau_{\nu}<t<0$, changes as a function of $\bm x$. Since the correlation scale of velocity gradients is also the viscous scale then we conclude that $l_{\nu}$ is the characteristic scale of variations of ${\hat n}\nabla\bm v{\hat n}(t, \bm x)$.

Exponential separation implies that trajectories that stay within the correlation length $l_{\nu}$ from each other during time interval $t$ must be initially separated by distance of order $l_{\nu}\exp(-\lambda_1 t)$ or smaller. In other words for $t\gg\tau_{\nu}$ the integrand of the spatial integral in
%PJK: The equation should appear before referred to
Eq.~(\ref{space}) varies over a characteristic scale $l_{\nu}\exp(-\lambda_1 t)$. Therefore effectively the space average is carried over roughly $(L/l_{\nu}\exp(-\lambda_1 t))^3$ independent random variables. Here $L$ is the system size which is at least the integral scale so that $(L/l_{\nu})^3\gtrsim Re^{9/4}$. The number of independent random variables grows exponentially and quickly gets so large that the law of large numbers applies. The RHS of
%PJK: The equation should appear before referred to
Eq.~(\ref{space}) becomes then time-independent and provides $\mu$.

The inclusion of intermittency is necessary because, as we saw previously, $\mu$ is determined by quiescent eddies whose correlation time is proportional to a power of $Re$. As a result the RHS of
%PJK: The equation should appear before referred to
Eq.~(\ref{space}) will only become time independent at $t\sim \tau_{\nu}Re^{\rho}$ with some $\rho$ numerically close to $\kappa\sim 0.3$ above. It is possible to rewrite Eq.~(\ref{space}) in the form that demonstrates
that $\mu$ is time average of time-integrated spatial correlation function
\begin{eqnarray}&&\!\!\!\!\!\!\!\!\!\!\!\!\!\!
\mu\!=\! 2
\int_0^t \frac{ dt_1}{t}\int_{t_1}^t dt_2\int  d\bm x' \left({\hat n}\nabla\bm v{\hat n}(t_1, \bm x')\!-\!\lambda_1\right)
\nonumber\\&& \!\!\!\!\!\!\!\!\!\!\!\!\!\!\times
\left({\hat n}\nabla\bm v{\hat n}(t_2, \bm q(t_2| t_1, \bm x))\!-\!\lambda_1\right),\label{space1}
\end{eqnarray}
where $\bm x'=\bm q(t_1, \bm x)$ in Eq.~(\ref{space}).  We introduced Lagrangian trajectories that depend on initial time
\begin{eqnarray}&&\!\!\!\!\!\!\!\!\!\!\!\!\!\!
\partial_t \bm q(t| t', \bm x)=\bm v(\bm q(t| t', \bm x), t),\ \ \bm q(t=t'| t', \bm x)=\bm x, \label{lt}
\end{eqnarray}
cf. Eq.~(\ref{ts}). Similar considerations can be made for higher order cumulants.

{\bf Exponent as series in powers of $Re$}---Phenomenology of turbulence \cite{frisch} implies that the cumulants behave as powers of $Re$ so that $\Delta_n$ in Eq.~(\ref{cumulants}) obey
\begin{eqnarray}&&\!\!\!\!\!\!\!\!\!\!\!\!\!\!\!\!\!\!
\tau_{\nu}\Delta_n=c_n Re^{\beta_n}, \label{rgy}
\end{eqnarray}
with a certain function $\beta_n$ and dimensionless functions $c_n$ that are either constants or depend on $Re$ slower than a power-law. The previously considered case of $\mu$ with $n=2$ is special. The equal time correlation function, obtained by setting $t_1=t_2$ in the integrand of Eq.~(\ref{space1}), is given by the square of velocity gradients that is not influenced by intermittency and whose average is of order $\epsilon/\nu$.
The non-trivial $Re-$dependence of $\mu$ comes from the size of the effective integration range in $t_2-t_1$ variable, the correlation time. In contrast, for $n\geq 3$ already the equal-time correlation functions in $\Delta_n$ are
influenced by intermittency and have non-trivial $Re-$dependence. These functions include moments of velocity gradients of order higher than two, that are due to intermittent bursts and are determined by velocity gradients larger than $\tau_{\nu}^{-1}$ by a power of $Re$, cf. Eq.~(\ref{phso}). %For instance in the frame of the multifractal model \cite{frisch}, the equal time-correlation function of order $n$ would be determined by rare bursts
%with probability proportional to $Re^{-\delta_2}$ with positive $\delta_2$. These events are relevant, despite the probability's smallness, because the associated velocity gradient magnitude $s_l$ is larger than $\tau_{\nu}^{-1}$ by $Re^{\delta_1}$ with $\delta_1>0$. Thus, whereas $\mu$ is formed by moderate velocity gradients with long correlations times, $\Delta_{n\geq 3}$ could be formed by large gradients.

%In fact phenomenology
%PJK: comma
In the frame of the traditional phenomenology of turbulence \cite{frisch}, however, the correlation time of gradients $s_l$ is $s_l^{-1}$, cf. the discussion of lifetime of the stretching rate above. Thus large gradients have small correlation times. This implies that, despite that equal time-correlation functions in $\Delta_n$ are formed by events with very large gradients, the contribution of these events into the time integral defining $\Delta_n$ could be negligible. This is because time integration approximately multiplies all gradients $s_l$ in $\Delta_n$, but one, by small correlation time $1/s_l$. %Thus, after time integration, the gradients $s_l$ contribute into the cumulant in Eq.~(\ref{rgy}) only $\tau_{\nu}^{-1} Re^{\delta_1-\delta_2}$. This estimate is obtained as product of characteristic value $\tau_{\nu}^{-1} Re^{\delta_1}$  of the single gradient remaining after time integration and the gradient's probability $\sim Re^{-\delta_2}$. The resulting contribution is negligible if $\delta_1-\delta_2<0$ and, if $\delta_1-\delta_2>0$ it can still be negligible in comparison with the contribution of moderate gradients with long correlation times.
%Thus, in fact, $\Delta_{n\geq 3}$ could be formed by moderate long-correlated velocity gradients.

%The above consideration demonstrates that if the lifetime of large gradients in turbulence can be estimated as the inverse gradient, and yet moderate gradients' lifetime can be much larger than the inverse gradient, then $\Delta_{n\geq 3}$ will be determined by the moderate gradients.
%PJK: Trying to reduce repetition of 'gradient'
The above consideration demonstrates that if the lifetime of large
%gradients in turbulence can be estimated as the inverse gradient, and
%PJK: maybe 'as their inverse' is enough
gradients in turbulence can be estimated as their inverse, and
%yet moderate gradients' lifetime can be much larger than the inverse
%gradient, then $\Delta_{n\geq 3}$ will be determined by the moderate
%PJK: maybe 'as their inverse' is enough
yet moderate gradients' lifetime can be much larger than their
inverse, then $\Delta_{n\geq 3}$ can be determined by the moderate
gradients.
%PJK
At the same time, it is plausible \cite{mv} that long-living vortices with large vorticity are associated in the turbulent flow with large strains (which are more relevant than vorticity regions since the trajectories separate predominantly in strain regions) whose correlation time exceeds the inverse strain by a power of $Re$, cf. \cite{bud}. Thus further studies are necessary to determine which events determine $\Delta_n$ and ultimately $\gamma(k)$ given by
\begin{eqnarray}&&\!\!\!\!\!\!\!\!\!\!\!\!\!\!
\gamma(k)\tau_{\nu}=c_1  Re^{-\delta}k+\frac{c_2Re^{\kappa} k^2}{2}+\sum_{n=3}^{\infty}\frac{c_nRe^{\beta_n}k^n}{n!}. \label{sq}
\end{eqnarray}
The lack of knowledge of how $c_n$ and $\beta_n$ depend on $n$ does not allow to fix the functional form of the dependence of $\gamma(k)$ on $Re$. It is plausible that in the limit $Re\to\infty$ (which could require $Re$ higher than those in most applications) one term determines the whole sum and then $\gamma(k)$ also obeys a power-law dependence on $Re$. (We remark that truncation of the series for $\gamma(k)$ does not contradict Pawula's theorem \cite{fk}.) This assumption is made stronger by the observations of \cite{mohan} for the Lyapunov exponent of turbulence $\lambda_1^v$, considered in more detail later. We demonstrate below that $\lambda_1^v$ is similar to $\gamma(2)$. Since the former is observed to obey power-law dependence in
%$Re$ then similar
%PJK: added a comma and 'a'
$Re$, then a similar
dependence would hold for $\gamma(2)$ (and then also for other $\gamma(k)$).
%Other indication
%PJK: Other => Further
Further indication
of power-law behavior is obtained by considering high $Re$ where the first term in the series in Eq.~(\ref{sq}) is negligible. There we have $\gamma(k)\approx c_2Re^{\kappa} k^2/2$ for $k$ small (yet not as small as $\lesssim Re^{-\kappa-\delta}$). Then, if we can assume that e.g. for $k=0.1$ the last equality holds uniformly in $Re$ then the monotonic increase of $\gamma(k)$ with $k$ would imply that $\gamma(2)$ is bounded from below by $Re^{\kappa}$ times a (small) constant. The last very plausible demonstration that $\gamma(k)$ must obey power-law behavior in $Re$ comes from the large deviations theory considered below.

\subsection{Large deviations and hyper-intermittency} \label{hi}

The dependence of the moments of $r(t)$ on time, described by Eq.~(\ref{iwn}) implies that the probability density function of $\rho(t, \bm x)\equiv t^{-1}\ln (r(t, \bm x)/r_0)$ is described by the large deviations theory \cite{frisch,fk}
\begin{eqnarray}&&\!\!\!\!\!\!\!\!\!\!\!\!\!\!
P(\rho, t)\sim  e^{-tS(\rho)},\ \ \langle r^k(t)\rangle_s\sim \int  e^{t(k\rho-S(\rho))}d\rho, \label{pf}
\end{eqnarray}
where $S(\rho)$ is the so-called large deviations function. At large times the integral in Eq.~(\ref{pf}) is determined by the maximum of the exponent demonstrating that $\gamma(k)$ and $S(\rho)$ form a Legendre transform pair, $\gamma(k)=\max_{\rho}[k\rho-S(\rho)]$. It is readily seen from the properties of the Legendre transformation that $S(\rho)$ is a convex non-negative function that has a unique minimum of zero at $\rho=\lambda_1$. Thus $S(\rho)$ describes the volume fraction (recall that the averages are spatial) of regions where the asymptotic equality $\ln r(\bm x, t)/t\approx \lambda_1$ is violated significantly at however large $t$. The volume of these regions decays exponentially thus leading to $\lim_{t\to\infty} P(\rho, t)=\delta(\rho-\lambda_1)$, implied by $\exp(-tS(\rho))$ form. This is equivalent to the statement that $\lambda_1(\bm x)$ in Eq.~(\ref{sv}) equals $\lambda_1$  for almost all $\bm x$.

Intermittency implies that $\gamma(k)$ is
%non-linear and
%PJK: added a comma and 'that'
non-linear, and that
the maximum of $k\rho-S(\rho)$ is attained at some $\rho=q(k)$ which is different from $\lambda_1$. This signifies that extremely rare regions of space, whose volume fraction $\sim \exp\left(-tS(q(k))\right)$ is exponentially small at large times, determine $\langle r^k\rangle_s$. This happens because $r(t, \bm x)\sim \exp(q(k) t)$ grows there anomalously fast, see above and e. g. detailed discussion in \cite{fk}. It is seen from the definition $\rho(t, \bm x)\equiv t^{-1}\int_0^t {\hat n}\nabla \bm v {\hat n} dt'$ that qualitatively $q(k)$ is the
%time-average value of velocity gradients that determine
%PJK: time-averaged ... determines
time-averaged value of velocity gradients that determines
the $k-$th moment of the distance. It seems plausible that this value has a power-law behavior in $Re$ similarly to the gradients that determine the usual single-time moments of velocity gradients. Then, since $kq(k)\sim S(q(k))$ is implied by maximization of $k\rho-S(\rho)$, we find
\begin{eqnarray}&&\!\!\!\!\!\!\!\!\!\!\!\!\!\!
q(k)\propto S(q(k))\propto Re^{\zeta(k)},\ \ \gamma(k)\tau_{\nu}=b_k Re^{\zeta(k)}, \label{gr}
\end{eqnarray}
%where $b_k$ are constants or slow functions of $Re$ and $\zeta(k)$ is a non-trivial function.
%PJK: slow => weak, added comma
where $b_k$ are constants or weak functions of $Re$, and $\zeta(k)$ is a non-trivial function.

{\bf Hyper-intermittency}---We provide quantitative description of the hyper-intermittency considered in the Introduction.  The above equalities tell that the volume fraction of spatial regions that determine $\langle r^k\rangle_s$ behaves as $\sim \exp\left(-t{\tilde b}_k Re^{\zeta(k)}\right)$ with quasi-constant functions of the Reynolds number ${\tilde b}_k$ (here and below we call functions that are either constants or depend on $Re$ slower than a power-law "quasi-constants" as their role is no different from constants. These functions depend on $k$ though). The regions are exponentially rare in both time and $Re$. Similarly, the growth of the moment of the distance is exponential both in time and $Re$. The physics of this growth is that it occurs due to gradients that are larger than a typical value $\tau_{\nu}^{-1}$ by a power of $Re$ and are also preserved on average during very long time. This is more  described above.

Jensen's inequality $ \exp\left(k\int_0^t\!\! \left\langle {\hat n}\nabla \bm v {\hat n}\right\rangle dt' \right)\leq \left\langle \exp\left(k\int_0^t\!\! {\hat n}\nabla \bm v {\hat n}dt' \right)\right\rangle$ implies that $\gamma(k)\geq k\lambda_1$. Due to intermittency this inequality is strict. The inequality $\gamma(k)/(k\lambda_1)>1$ is well-known and holds also for Gaussian velocities \cite{reviewt}. However, this intermittency is non-parameteric and $\gamma(k)/(k\lambda_1)\sim 1$ for moderate $k$. Here we stress that for turbulence the deviation from the self-similar growth of distances that is described by $\gamma(k)=k\lambda_1$ is parameteric, i.e. $\gamma(k)/(k\lambda_1)$ is a power of $Re$ with a $k$-dependent exponent.

\subsection{Strong non-parabolicity of $\gamma(k)$}

The results above imply that the very usual quadratic approximation to $\gamma(k)$, e.g. arising in the so-called Kraichnan model \cite{review}, is inconsistent with intermittency. Indeed, $\gamma(k)$ obeys the constraint $\gamma(-3)=0$,  see \cite{zeld,fk}. Together with $\gamma'(0)=\lambda_1$ and $\gamma(0)=0$ this constrains the quadratic approximation to the form $\gamma(k)=\lambda_1 k(k+3)/3$. This would imply that $\mu$ has the same dependence on $Re$ as $\lambda_1$ producing inconsistency. The deviation from parabolicity grows indefinitely with $Re$ since, neglecting the weak dependence of $\lambda_1\tau_{\nu}$ on $Re$, we have $\mu/\lambda_1\sim Re^{\kappa}$. In contrast, the quartic approximation to $\gamma(k)$, that was demonstrated in \cite{fk} to agree with the available observations, does not have this deficiency.

It could also be thought that quadratic approximation provides a lower bound for $\gamma(k)$ and intermittency implies faster than parabolic (Gaussian) growth,
\begin{eqnarray}&&\!\!\!\!\!\!\!\!\!\!\!\!\!\!
\gamma(k)\geq k\lambda_1+\frac{\mu k^2}{2},
\end{eqnarray}
cf. Eq.~(\ref{cumulants}). However, this inequality cannot be true at any $k>0$. Indeed, the skewness of velocity derivatives is negative which implies that $\Delta_3$ is most probably negative. Then, Eq.~(\ref{cumulants}) implies that the inequality above does not hold for small $k$. Yet, for large $k$ the inequality must be true because the decay of $P(\rho, t)$ is indeed slower than Gaussian, see \cite{fk} and references therein. Thus the above inequality is true for $k$ larger than some $k_0$. Since $\Delta_4$ is positive and, belonging to higher order moment of velocity derivatives, it grows faster with $Re$ than $\Delta_3$, then $k_0\sim 1$ is probable. The detailed study is left for future work.

\subsection{Resume and numerical study of $\gamma(k)$}

The main contribution of this section is providing an approach to the study of the $Re-$dependence of the generalized Lyapunov exponent. This approach relies on the observation that the cumulants have a power law dependence on $Re$ and represents $\gamma(k)$ as a series in the cumulants. This seems to be of use not only theoretically. The numerical studies of $\gamma(k)$ depends on the accumulation of statistically significant pool of data that contains stretching histories whose fraction decays with time (hyper)exponentially. Our study indicates that the higher $Re$ is the more demanding this becomes. At the same time the well-developed accurate procedures that are used for measuring the $Re-$dependence of equal time moments of velocity gradients may be employed for the numerical study of the cumulants. As of today, the dependence was measured (with varying accuracy) for moments up to $12-$th order, see the references in \cite{lee}.
%Thus hopefully the $Re-$dependence could
%PJK: expanded somewhat
Thus the $Re-$dependence can be obtained for $\Delta_n$ with
  $n\leq 12$ for the highest resolution simulations to date.
%PJK
%Summation of the first twelve terms in Eq.~(\ref{cumulants}) then would give $\gamma(k)$ for a significant range of $k$.
%PJK: avoid conditional
Summation of the first twelve terms in Eq.~(\ref{cumulants})
will give $\gamma(k)$ for a significant range of $k$.

\section{Infinite-dimensional Lyapunov exponent: magnetic field} \label{mag}

The results of the previous section are manifested most remarkably by considering growth of small fluctuations of magnetic field $\bm B$ in turbulent flows of ideally conducting fluids. The implications are immediate in the Lagrangian frame and can be transferred to the Eulerian frame that is more relevant in this case. The magnetic field obeys the induction equation
\begin{eqnarray}&&\!\!\!\!\!\!\!\!\!\!\!\!\!\!
\partial_t\bm B+(\bm v\cdot\nabla)\bm B=(\bm B\cdot\nabla)\bm v, \label{induction}
\end{eqnarray}
where the flow $\bm v$ is considered to obey the usual Navier-Stokes equations. Thus we study the kinematic dynamo where $\bm B$ is considered to be small so that the Lorentz force in the momentum equation is negligible. The magnetic field in the particle's frame obeys equation identical to Eq.~(\ref{s})
\begin{eqnarray}&&\!\!\!\!\!\!\!\!\!\!\!\!\!\!
\frac{d}{dt} \bm B(t, \bm q(t, \bm x))=(\bm B(t, \bm q(t, \bm x))\cdot \nabla)\bm v. \label{fa}
\end{eqnarray}
This equation describes the well-known fact that at zero magnetic diffusivity magnetic field lines are frozen in the fluid and behave as its infinitesimal line elements \cite{ll8}.
We find immediately that
\begin{eqnarray}&&\!\!\!\!\!\!\!\!\!\!\!\!\!\!
\lim_{t\to\infty}\!\frac{\ln B(t, \bm q(t, \bm x))}{t}\!=\!\lambda_1(\bm x),\ \ \langle B^k(t, \bm q(t, \bm x))\rangle_s\! \sim \!e^{\gamma(k)t},\nonumber
\end{eqnarray}
where we assume that $B(t=0, \bm x)$ is bounded and omit the initial condition factor in the argument of the logarithm for conciseness. Thus for almost all $\bm x$ the asymptotic logarithmic growth rate of $B(\bm q(t, \bm x), t)$ is $\lambda_1$. We also observe that $\bm B(t, \bm x)=B{\hat n}(t, \bm x)$ at $t\gg \tau_{\nu}$. Indeed, the relaxation to the major stretching direction occurs also for events that involve large deviations of the type described in the previous Section i.e., holds for all relevant events \cite{fp}.
%(the relaxation may not occur only for atypical local evolutions where infinitesimal spheres of the fluid are transformed by the flow into ellipsoids whose largest axes are equal. These events have no relevance here, cf. \cite{fb} {\color{em}{either make comment in parenthesis clearer or delete it altogether}}).

Incompressibility implies that there is no difference between the Lagrangian and Eulerian averages (in compressible case things are quite different \cite{review,arxiv}). Thus for the Eulerian average $\langle B^k\rangle_s$ we have
\begin{eqnarray}&&\!\!\!\!\!\!\!\!\!\!\!\!\!\!
\langle B^k(t)\rangle_s\!\equiv\! \!\!\int\!\! B^k(t, \bm x') d\bm x'\!=\!\!\!\int \!\!B^k(t, \bm q(t, \bm x)) d\bm x\!\sim\! e^{\gamma(k)t},\label{sf}
\end{eqnarray}
where $\bm x'=\bm q(t, \bm x)$. This equality implies also that the asymptotic logarithmic growth rate at a fixed point in space is $\lambda_1$ so that
\begin{eqnarray}&&\!\!\!\!\!\!\!\!\!\!\!\!\!\!
\lim_{t\to\infty}\!\frac{\ln B(t, \bm x)}{t}\!=\!\lambda_1,\label{ps}
\end{eqnarray}
for almost all $\bm x$. To demonstrate the above we observe that
\begin{eqnarray}&&\!\!\!\!\!\!\!\!\!\!\!\!\!\!
\ln\frac{B(t, \bm x)}{B(0, \bm q(0| t, \bm x))}\!=\!\int_0^t dt' \left[{\hat n}\nabla \bm v {\hat n}\right](t', \bm q(t'| t, \bm x)), \label{sa}
\end{eqnarray}
which is a form of Eq.~(\ref{log}) where we use $\bm q(t'| t, \bm x)$ introduced in Eq.~(\ref{lt}) with $t'<t$. We assume $t\gg \tau_{\nu}$ in Eq.~(\ref{sa}) which allows to neglect the transients to equality $\bm B(t, \bm x)/B(t, \bm x)\approx {\hat n}(t, \bm x)$. We find dividing Eq.~(\ref{sa}) by $t$ and letting $t\to\infty$ that by the multiplicative ergodic theorem the limit is constant for almost all $\bm x$. This constant can be found by
%space
%PJK: spatial
spatial
averaging of Eq.~(\ref{sa}). Transformation of the integration variable from $\bm x$ to $\bm q(t'| t, \bm x)$ reproduces the form of $\lambda_1$ given by Eq.~(\ref{las}) yielding Eq.~(\ref{ps}). The same conclusion is implied by $\left\langle \ln B\right\rangle/t=\lambda_1$
obtained by differentiating Eq.~(\ref{sf}), setting $k=0$ and using $\gamma'(0)=\lambda_1$. Yet another proof will be given below.

%We find differentiating over $k$ and setting $k=0$ that
%\begin{eqnarray}&&\!\!\!\!\!\!\!\!\!\!\!\!\!\!
%\lim_{t\to\infty} \frac{\langle \ln B(t, \bm x)\rangle}{t}=\lambda_1,
%\end{eqnarray}
%which difference from the previous result is that $\bm B$ is considered at a fixed point in space and there is averaging. However in fact $t^{-1}\ln B(t, \bm x)$ converges to the same constant for almost all $\bm x$ as can be seen by applying the ergodic theorem to
%\begin{eqnarray}&&\!\!\!\!\!\!\!\!\!\!\!\!\!\!
%\lim_{t\to\infty}\!\frac{\ln B(t, \bm x)}{t}=\lim_{t\to\infty}\int_0^t {\hat n}\nabla \bm v {\hat n}\frac{dt'}{t}.
%\end{eqnarray}
%This time ${\hat n}\nabla \bm v {\hat n}$ is evaluated on the Lagrangian trajectory passing through $\bm x$ at time $t$ and not through $\bm x$ at $t=0$, cf. Eq.~(\ref{sv}). The constant limit is then equal to the spatial average $\langle \ln B(, \bm x)\rangle$ so that
%\begin{eqnarray}&&\!\!\!\!\!\!\!\!\!\!\!\!\!\!
%\lim_{t\to\infty} \frac{\ln B(t, \bm x)}{t}=\lambda_1,\label{limit}
%\end{eqnarray}
%that holds for all $\bm x$ with possible exception of $\bm x$ with zero total volume.

\subsection{Non-uniform convergence to $\lambda_1$}

The description of the spatial distribution of the exponentially growing field $\bm B(t, \bm x)$ can be transferred from that of $r(t, \bm x)$ and is briefly provided here. The asymptotic in time growth of the magnetic field at the same rate almost everywhere in no way implies that there is ever a time where $\ln B(\bm x, t)/t\approx \lambda_1$ is a good approximation uniformly in space. For instance, no matter how large $t$ is,  the estimate $\int B^2(t, \bm x)d\bm x\sim \exp(2\lambda_1 t)$ does not provide a viable approximation for the magnetic energy whose true value is exponentially larger and is given by $\exp(\gamma(2)t)$. The reason is intermittency. That causes an exponentially decaying in time fraction of the total volume, where the growth rate of the magnetic field deviates significantly from $\lambda _1$, to carry most of the energy. The time $t^*(\bm x)$ starting from which $\ln B(t, \bm x)/t\approx \lambda_1$ holds, strongly depends on $\bm x$. We introduce the exponential growth field $\theta$, which is similar to $\rho$ above, by
\begin{eqnarray}&&\!\!\!\!\!\!\!\!\!\!\!\!\!\!
B(t, \bm x)=\exp\left(\theta(t, \bm x)t\right), \ \ \theta(t, \bm x)\equiv \frac{\ln B(t, \bm x)}{t}.
\end{eqnarray}
Thus $\theta(t, \bm x)$ is the growth exponent of the magnetic field at the fixed point in space $\bm x$. Its spatial fluctuations are described by the same large deviations theory as $\rho$
\begin{eqnarray}&&\!\!\!\!\!\!\!\!\!\!\!\!\!\!
%P(\theta, t)\sim  e^{-tS(\theta)},\ \ \langle B^k(t)\rangle\sim \int  e^{t(k\theta-S(\theta)}d\theta, \label{pdf}
%PJK: added missing bracket
  P(\theta, t)\sim  e^{-tS(\theta)},\ \ \langle B^k(t)\rangle\sim \int  e^{t(k\theta-S(\theta)\red{)}}d\theta, \label{pdf}
\end{eqnarray}
where $S$ is the same function that we introduced in the previous section due to equality of Eulerian and Lagrangian averages. This probability density function (PDF) implies as previously that $\lim_{t\to\infty}P(\theta, t)=\delta(\theta-\lambda_1)$ providing another proof of Eq.~(\ref{ps}). Quadratic expansion of $S(\theta)$ near the minimum at $\lambda_1$ produces the central limit theorem \cite{fb,gaw,review}. The probability distribution function of $(\ln B(\bm q(t, \bm x), t)-\lambda_1 t)/\sqrt{t}$ (and thus also of $(\ln B(\bm x, t)-\lambda_1 t)/\sqrt{t}$) becomes Gaussian in the infinite time limit with dispersion $\mu$ (similar conclusion holds for $r(t, \bm x)$).

The most relevant of the exponents $\gamma(k)$, described by Eq.~(\ref{gr}), in this case is $\gamma(2)$. It describes the growth rate of the magnetic energy and thus determines when the neglect of the magnetic component of the full energy of the conducting fluid becomes inconsistent. The previous results imply that the energy is concentrated in exponentially rare spatial regions whose volume fraction decays as $\exp(-S(q(2))t)$ where $S(q(2))\propto Re^{\zeta_2}$. This hyper-intermittency implies that accurate measurement of the growing magnetic energy is an exceedingly difficult task at high $Re$. The estimate of the growth rate of energy by $\tau_{\nu}^{-1}$ fails completely at large $Re$. Similar conclusions hold for other moments of the magnetic field.

\subsection{Field view}

The studied case provides a bridge between finite and infinite-dimensional systems that will be useful below. Indeed the induction equation, given by Eq.~(\ref{induction}), written in the form $\partial_t\bm B={\hat L}(t)\bm B$ where ${\hat L}(t)$ is a linear operator, can be considered as infinite-dimensional generalization of the equation on separation written in the form ${\dot r_i}=L_{ik}(t)r_k$ with $L_{ik}\equiv \nabla_kv_i$, see Eq.~(\ref{s}). We observe that trivially, in a finite-dimensional system we can use any of $l_k$ norms in the definition of the Lyapunov exponents that is
\begin{eqnarray}&&\!\!\!\!\!\!\!\!\!\!\!\!\!\!
\lambda_1\!=\!\lim_{t\to\infty}\frac{\ln \left(||\bm r(t)||_k/||\bm r(0)||_k\right)}{t},\nonumber\\&&\!\!\!\!\!\!\!\!\!\!\!\!\!\! ||\bm r(t)||_k\!=\!\left(\sum_{i=1}^3|r_i(t)|^k\right)^{1/k}.
\end{eqnarray}
It is readily seen that $\lambda_1$ defined by the above equation is independent of $k$, since the limit implies that the maximal component of $\bm r$ grows at the rate $\lambda_1$, see the Introduction. In contrast, the infinite-dimensional counterpart of the above equation,
\begin{eqnarray}&&\!\!\!\!\!\!\!\!\!\!\!\!\!\!
\lim_{t\to\infty}\frac{\ln  \left(||\bm B(t)||_k/||\bm B(0)||_k\right)}{t}\!=\!\frac{\gamma(k)}{k},\ \ \\&&\!\!\!\!\!\!\!\!\!\!\!\!\!\!
||\bm B(t)||_k\!\equiv \!\left(\int\!\! B^k(t, \bm x)d\bm x\right)^{1/k},\nonumber
\end{eqnarray}
depends on $k$.
%This gives more grounding
%PJK: reworded
This gives a more solid foundation
%PJK
to the name "generalized Lyapunov exponent" since $\gamma(k)/k$ is proportional to the true infinite-dimensional Lyapunov exponent defined with the help of the $L_k$ norm. Thus our assumption that $\gamma(k)$ is independent of realization of velocity is equivalent to the assumption of realization-independence of infinite-dimensional Lyapunov exponents defined with the corresponding norm.

\subsection{Hyper-intermittency law}

Finally we give another outlook at the results of this section which will be useful below. It is found by differentiating Eq.~(\ref{induction}) that
\begin{eqnarray}&&\!\!\!\!\!\!\!\!\!\!\!\!\!\!
\frac{d}{dt}\int B^{2k}(t, \bm x)d\bm x=2k \int B^{2k-2} B_iB_k\nabla_i v_k d\bm x,  \label{2lk}
\end{eqnarray}
where we assume that the boundary term can be neglected or is zero as in the case of periodic or zero normal velocity boundary conditions. We find, recalling the alignment of $\bm B$ with the major stretching direction, that
\begin{eqnarray}&&\!\!\!\!\!\!\!\!\!\!\!\!\!\!
\frac{d}{dt}\ln \int B^{2k}(t, \bm x)d\bm x=2k \left\langle {\hat n}\nabla \bm v {\hat n}\right\rangle_{2k},  \label{k}
\end{eqnarray}
where we defined the average weighted with normalized measure $B^{2k}(t, \bm x)/ \int B^{2k}d\bm x$ by
\begin{eqnarray}&&\!\!\!\!\!\!\!\!\!\!\!\!\!\!
\left\langle {\hat n}\nabla \bm v {\hat n}\right\rangle_{2k}\equiv
\frac{\int B^{2k} {\hat n}\nabla \bm v {\hat n}d\bm x}{\int B^{2k}(t, \bm x)d\bm x}.
\end{eqnarray}
We obtain from the above by using Eq.~(\ref{gr}) and definition of $\gamma(k)$ that
\begin{eqnarray}&&\!\!\!\!\!\!\!\!\!\!\!\!\!\!
\left\langle {\hat n}\nabla \bm v {\hat n}\right\rangle_k=\frac{\gamma(k)}{k}=\frac{b_{k} Re^{\zeta(k)}}{k\tau_{\nu}}.
\end{eqnarray}
The last identities provide robust description of the hyper-intermittency of the growth. The normalized measure proportional to $B^{2k}$ is concentrated in the regions where the gradients are larger than inverse Kolmogorov time by a power of $Re$. Conversely, the $2k-$th moment of the field, which includes the energy, grows in regions where due to intermittency the gradients are larger than $\tau_{\nu}^{-1}$ by $Re^{\zeta(k)}$.

\section{Finite magnetic diffusivity}

In this section we demonstrate that the $Re-$dependence of the growth of the magnetic field in an ideally conducting
%fluids,
%PJK: singular
fluid,
derived previously, generalizes qualitatively to the case of finite magnetic diffusivity $\eta$. The main reason is that since the field growth at $\eta\ll \nu$ occurs mainly on spatial scales smaller than the Kolmogorov scale and larger than the diffusive scale \cite{falk}, it may be described with the help of finite-time Lyapunov exponents \cite{gaw,reviewt,fb}. The latter do not differ qualitatively from the previously considered $\rho$ variable, see \cite{fk} for relation between the exponents. Here we refer to previous works on the case of small diffusivity, reconsidering their formulas from the viewpoint of their Reynolds number dependence, which was not done previously.

The ideal fluid approximation studied in the previous section has a limited domain of validity. It is valid when the magnetic diffusivity $\eta$ that appears in the full induction equation \cite{ll8}
\begin{eqnarray}&&\!\!\!\!\!\!\!\!\!\!\!\!\!\!
\partial_t\bm B+(\bm v\cdot\nabla)\bm B=(\bm B\cdot\nabla)\bm v+\eta\nabla^2\bm B,
\end{eqnarray}
is much smaller than $\nu$, cf. Eq.~(\ref{induction}). This condition is usually referred to as the condition of large magnetic Prandtl number $Pr\equiv \nu/\eta$. If $Pr\gg 1$ then there is a finite interval of time during which the smallest spatial scale of the magnetic field is not too small and the diffusivity term is negligible everywhere. During such interval the assumption of ideal fluid applies. However, the mixing property of turbulence generates increasingly smaller scales unless the diffusivity term with higher order derivative becomes relevant for any value of $\eta$ regardless of how small it is \cite{ll8}. The resulting impact of $\eta$ on the growth exponents
\begin{eqnarray}&&\!\!\!\!\!\!\!\!\!\!\!\!\!\!
\gamma^B(k, \eta)\equiv \lim_{t\to\infty} \frac{1}{t} \ln\left(\frac{\int\!\! B^k(t, \bm x)d\bm x}{\int\!\! B^k(0, \bm x)d\bm x}\right), \label{gt}
\end{eqnarray}
is thus finite for $\eta>0$. We demonstrated previously that at $\eta=0$ it is highly plausible that the limits above exist and do not depend on the flow
%realization and the
%PJK: and => or
realization or the
initial conditions $\bm B(0, \bm x)$. It seems that finite magnetic diffusivity, that enhances the mixing of $\bm B$, seemingly can only strengthen the independence of the limit of the flow realization. Thus we assume below that $\gamma^B(k, \eta)$ are well-defined realization-independent properties.

%The realization-independence can be actually proved
%PJK: reordered
The realization-independence can actually be proven
by using the formula for $B^2$ obtained in \cite{falk} in the case of $Pr\gg 1$ and not too large times, see below. %We remark that the formula of \cite{falk} provides $B^2$ averaged over initial conditions $\bm B(t=0, \bm x)$  which are assumed to have zero spatial average (it can be seen that the average field evolves as in $\eta=0$ case). This does not allow to show that $\gamma^B(k)$ is independent of the initial conditions which can however be shown otherwise with details published elsewhere.
Thus \cite{falk} provide $B^2$ as exponential of linear combinations of finite-time Lyapunov exponents. These exponents are similar to the $\rho$ variable above, which is also a type of finite-time Lyapunov exponent, see definition and description in \cite{fb,gaw,reviewt,fk}. The application of the cumulant expansion theorem to the spatial average then demonstrates the existence and realization-independence of $\gamma^B(k, \eta)$ quite similarly to the demonstration for $\gamma(k)$.

The formula for $\gamma^B(k, \eta)$ was obtained in \cite{falk}. This work can be used to demonstrate that despite that $\gamma^B(k, \eta)$ is discontinuous at $\eta=0$ the nature of the $Re-$dependence of the growth is unchanged. To see that, we observe that as is demonstrated in \cite{falk},  the point-wise growth of the magnetic field is given by
\begin{eqnarray}&&\!\!\!\!\!\!\!\!\!\!\!\!\!\!
\lim_{t\to\infty}\!\frac{\ln B(t, \bm x)}{t}\!=\!\min\left(\frac{\lambda_1\!-\!\lambda_2}{2}, \frac{\lambda_2\!-\!\lambda_3}{2}\right),\ \ Pr\!\gg\! 1, \label{c1}
\end{eqnarray}
which in the limit $\eta \to 0$ is clearly different than the corresponding value at $\eta =0$, cf. Eq.~(\ref{ps}). Since the $Re-$dependence of the exponents is identical, see above, then we find
\begin{eqnarray}&&\!\!\!\!\!\!\!\!\!\!\!\!\!\!
\lim_{t\to\infty}\!\frac{\ln B(t, \bm x)}{t}\!=\!\frac{d\gamma_B}{dk}(k=0)=\frac{{\tilde c}_0}{\tau_{\nu}}Re^{-\delta},\label{c2}
\end{eqnarray}
where $\delta$ is the same exponent that appears in Eq.~(\ref{prediction}) and ${\tilde c}_0$ is quasi-constant. Thus, despite the discontinuity in the growth rate at $\eta =0$, its dependence on $Re$ is the same for $\eta \to 0$ and $\eta =0$ (the constant ${\tilde c}_0$ is smaller than its counterpart at $\eta=0$).

The $k-$th moment of the magnetic field can be found as the average of an exponent containing linear combinations of finite-time Lyapunov exponents \cite{falk}. The exponents obey statistics similar to that of $\rho$ above \cite{fb,reviewt}. The result of the averaging is therefore similar
\begin{eqnarray}&&\!\!\!\!\!\!\!\!\!\!\!\!\!\!
\gamma^B(k)\tau_{\nu}={\tilde b}_k Re^{{\tilde \zeta}(k)},\ \ \eta\to 0 \label{gh}
\end{eqnarray}
where the scaling exponents ${\tilde \zeta}(k)$ and quasi-constants ${\tilde b}_k$ generally differ from $\zeta(k)$ and $b_k$ in Eq.~(\ref{gr}) except for ${\tilde \zeta}(0)=\zeta(0)$. Generally, a strict inequality is realized in $\gamma^B(k, \eta\to 0)\leq  \gamma^B(k, \eta=0)$ implying ${\tilde \zeta}(k)<\zeta(k)$.

A possible difference between $\gamma^B(k)$ and $\gamma(k)$ is their analytic properties. The function $\gamma(k)$ has continuous derivatives and might well be analytic, see Eq.~(\ref{cumulants}). In contrast, $\gamma^B(k)$ might not
%by analytic.
%PJK: typo
be analytic.
Changes of regime producing discontinuities in the first derivative of $\gamma^B(k)$ are possible similarly to \cite{fb} where for some $k$ the average is determined by the boundary of the range of variations of the finite-time Lyapunov exponents and for some by the interior. The existence of points with discontinuous first derivative depends on the form of the large deviations function of the finite time Lyapunov exponents and its study is beyond our scope here. In Kraichnan model $\gamma^B(k)$ was found to be analytic \cite{falk}.

The properties of averaging over finite-time Lyapunov exponents are similar to those of averaging over $\rho$. Exponential growth of the moments $\left\langle B^k(t)\right\rangle\sim \exp(\gamma^B(k)t)$ implies that the PDF of $\theta\equiv \ln B/t$ still has the large deviations form given by Eq.~(\ref{pdf}). The large deviations function ${\tilde S}(\theta)$ is the Legendre transform of $\gamma^B(k)$ that has minimal value zero attained at the argument equal to the RHS of Eq.~(\ref{c1}). Thus Eq.~(\ref{c1}) holds with probability one which implies that it holds also independently of $\bm B(t=0)$ and realization of velocity. The volume fraction of the regions that determine a given moment is again exponentially small both in time and in the Reynolds number as in the case of $\eta=0$. The detailed derivation of above properties from \cite{falk} is beyond the scope of this paper and will be published elsewhere.

\subsection{Hyper-intermittency law at $\eta>0$}

We can reinterpret the above results in terms of a hyper-intermittency law similar to that in the previous section. We observe that
\begin{eqnarray}&&\!\!\!\!\!\!\!\!\!\!\!\!\!\!
\frac{d}{dt}\int B^{2k}(t, \bm x)d\bm x=2k \int d\bm x \left(  B_iB_k\nabla_iv_k+\eta B_i\nabla^2 B_i\right)
\nonumber\\&&\!\!\!\!\!\!\!\!\!\!\!\!\!\!
\times B^{2k-2}(t, \bm x) \sim 2k \int B^{2k-2} B_iB_k\nabla_iv_k  d\bm x,
\end{eqnarray}
cf. Eq.~(\ref{2lk}). The last asymptotic equality recognizes that the growth occurs when the magnetic lines' stretching and diffusivity terms are of the same order (otherwise either the growth exponents were as in $\eta=0$ case or the diffusivity term would dominate and stop the growth). We find that
\begin{eqnarray}&&\!\!\!\!\!\!\!\!\!\!\!\!\!\!
\frac{2k \int B^{2k-2} B_iB_k\nabla_iv_k  d\bm x}{\int B^{2k}(t, \bm x)d\bm x}\!\sim\! \gamma^B(k)\!=\!\frac{{\tilde b}_k}{\tau_{\nu}} Re^{{\tilde \zeta}(k)}. \label{grad}
\end{eqnarray}
This implies that qualitatively, as in $\eta=0$ case, the growth of the $2k-$th moment occurs due to gradients that are larger than $\tau_{\nu}^{-1}$ by a power of $Re$. The geometry of magnetic field configurations changes and the growth exponents are depleted by factor of order one \cite{falk}, yet the $Re-$dependence is similar to $\eta=0$ case.

The results of \cite{falk} hold for times large enough that the diffusivity is relevant at the smallest scale of $\bm B(t, \bm x)$, yet small enough that the linear size of the zone of dependence of $\bm B(t, \bm x)$ (that is the spatial region where changes of the initial conditions $\bm B(t=0)$ change the value of $\bm B(t, \bm x)$ appreciably) is much smaller than the viscous scale $l_{\nu}$. The last assumption is necessary in order to describe the velocity by linear profile as done in that study. Therefore \cite{falk} does not allow to study the limit $t\to\infty$ being confined, as can be seen, to time interval of order $\tau_{\nu}\ln Pr$ (correspondingly the infinite time limits in Eqs.~(\ref{c1})-(\ref{c2}) must be understood in asymptotic sense). Beyond these times the zone of dependence is larger than $l_{\nu}$. During the studied time interval of order $\tau_{\nu}\ln Pr$ exponentially growing moments of the magnetic field, including the energy, increase by a power of $Pr$.

The study of $t\gg \tau_{\nu}\ln Pr$, that ultimately
%fix
%PJK: fixes
fixes
$\gamma^B(k, \eta)$ in Eq.~(\ref{gt}), requires other techniques. For finding the energy exponent $\gamma^B(2)$ one can use the evolution equation for the spectrum of magnetic field \cite{gruzinov}, see also \cite{fl}. That equation is formulated in terms of finite-time Lyapunov exponents and hence the result of the solution will still have the form given by Eq.~(\ref{gh}), see also \cite{ha}. For other moments a different scheme needs to be developed where probably the most intriguing question is what comes instead of Eq.~(\ref{c1}). These studies are left for future work. Here we observe that there seem to be no reason
%for qualitative
%PJK: added 'a'
for a qualitative change of Eq.~(\ref{grad}) because the magnetic field is concentrated at a diffusive scale that is much smaller than the viscous scale and where the linear velocity approximation is valid.
Decrease of $Pr$ leads to increase of diffusive scale that becomes equal to the viscous scale at $Pr=1$. The asymptotic continuation then indicates that
\begin{eqnarray}&&\!\!\!\!\!\!\!\!\!\!\!\!\!\!
\gamma^B(k, \eta)\tau_{\nu}={\tilde b}_k(\eta) Re^{{\tilde \zeta}(k, \eta)},\ \ Pr\lesssim 1, %\label{gh}
\end{eqnarray}
with ${\tilde b}_k(\eta)$ and ${\tilde \zeta}(k, \eta)$ smooth functions of $\eta$. Thus  the growth of the magnetic field at $Pr=1$, as governed by,
\begin{eqnarray}&&\!\!\!\!\!\!\!\!\!\!\!\!\!\!
\partial_t\bm B+(\bm v\cdot\nabla)\bm B=(\bm B\cdot\nabla)\bm v+\nu\nabla^2\bm B, \label{Sas}
\end{eqnarray}
must still exhibit $Re-$dependent exponents. This agrees qualitatively with observations considered in the next section. The law given by Eq.~(\ref{gh}) is valid also at $Pr\ll 1$ with magnetic Reynolds number taking the place of $Re$ as will be reported in a successive publication.

To summarise, the limit of zero magnetic diffusivity is
%singular however not too singular.
%PJK: reformulated, Itzhak please check
singular, but the results at this limit are still useful.
%PJK
It seems that it preserves the power-law dependence on $Re$ of the growth exponents of the magnetic field moments that has been deduced for $\eta=0$. This can be proved rigorously in the intermediate time regime studied by \cite{falk}. This can also be proved at $t\to\infty$ for $\gamma_B(2)$ using the spectrum equation of \cite{gruzinov}. The full study is left for future work. Finally the results at $Pr\gg 1$ seem to extend to $Pr\sim 1$.

\section{$Re-$dependence of generalized Lyapunov exponents of turbulence}

In this section we introduce the generalized Lyapunov exponent of turbulence which describes the growth rate of different moments of the velocity perturbation. Probably the main contribution of this section to the current research is raising of the possibility
%that there exists a pointwise logarithmic growth rate
%PJK: ordering
that a pointwise logarithmic growth rate exists
%PJK
that holds almost everywhere at large times. This rate differs strongly from the quantity which is currently called the Lyapunov exponent of turbulence.

The generalized Lyapunov exponent of turbulence describes exponential growth of separation $\delta \bm v$ of two solutions of the Navier-Stokes (NS) equations given by Eqs.~(\ref{ns}). The evolution of $\delta \bm v$ is described by the linearized NS equations
\begin{eqnarray}&&\!\!\!\!\!\!\!\!\!\!\!\!\!\!
\partial_t\delta\bm v\!+\!(\bm v\cdot\nabla)\delta\bm v\!=\!-(\delta \bm v\cdot\nabla)\bm v\!-\!\nabla \delta p\!+\!\nu\nabla^2\delta \bm v,\label{das}
\end{eqnarray}
and the incompressibility condition $\nabla\cdot\delta\bm v=0$. The equation's difference from Eq.~(\ref{Sas}) consists of the different sign of the lines' stretching term and the presence of the pressure which is necessary to ensure solenoidality. These changes would not change the exponential growth of $\delta \bm v$ similar to $\bm B$. The issue is how we measure this growth. The usually used definition of the Lyapunov exponent of turbulence is
\begin{eqnarray}&&\!\!\!\!\!\!\!\!\!\!\!\!\!\!
\lambda^v\equiv \lim_{t\to\infty}\frac{1}{t}\ln\left(\frac{\left(\int (\delta v)^2(\bm x, t)d\bm x\right)^{1/2}}{\left(\int (\delta v)^2(\bm x, 0)d\bm x\right)^{1/2}}\right),\label{ad}
\end{eqnarray}
see e.g. \cite{ruelle,cr,aurell,mohan}. It is however obvious from the previous studies that the preference for the $L_2$ norm is not self-evident and a more versatile characteristics of the growth is provided by the generalized Lyapunov exponent $\gamma^v(p)$ defined in the Introduction
\begin{eqnarray}&&\!\!\!\!\!\!\!\!\!\!\!\!\!\!
\gamma^v(p)\equiv \lim_{t\to\infty}\frac{1}{t}\ln\left(\frac{\left(\int \left|\delta v\right|^p(t, \bm x)d\bm x\right)^{1/p}}{\left(\int \left|\delta v\right|^p(0, \bm x)d\bm x\right)^{1/p}}\right),
\end{eqnarray}
where $\lambda^v=\gamma^v(2)$. We see that for any $p$ we can interpret $\gamma^v(p)$ as "the" Lyapunov exponent of the NS equation should we decide to declare $L_p$ as "the" norm on the functional space. Thus $\gamma^v(p)$ are probably realization-independent quantities which is reinforced by their similarity with $\gamma^B(p, \eta=\nu)/p$ considered in the previous section. More generally there is qualitative similarity between
$\gamma(k)$, $\gamma^B(k)$ and $k\gamma^v(k)$. Thus $\lambda^v$ must behave as $\gamma(2)$ and has a
%power law
%PJK: hyphen used elsewhere
power-law type growth with $Re$ as indeed observed in \cite{mohan}. This fully explains the discrepancy of the observed behavior of $\lambda^v$ and that predicted in \cite{cr}, see the qualitative explanation in the Introduction.
%Indeed, the definition of $\gamma(2)$ gives rough estimate $\gamma(2)\sim t^{-1} \ln \left\langle \exp(2t/\tau_{\nu})\right\rangle$ where $\tau_{\nu}$ is a characteristic viscous time {\color{em}{explain how this result has been obtained}}. This is very different from $\left\langle \tau_{\nu}^{-1}\right\rangle$ which was used in \cite{cr} for estimating $\lambda_1^v$.

The above analogy would indicate the probable validity of
\begin{eqnarray}&&\!\!\!\!\!\!\!\!\!\!\!\!\!\!
\left\langle |\delta v|^k \right \rangle_s\sim \exp(\gamma^v(k) t),\ \
\gamma^v(k)\tau=d(k)Re^{\xi(k)},
\end{eqnarray}
where $d(k)$ are quasi-constants. This form implies, similarly to our previous studies, that the PDF of $\phi(t, \bm x)\equiv t^{-1}\ln |\delta \bm v(t, \bm x)|$ has large deviations form $\sim \exp\left(-tH(\phi)\right)$. Here $H$ is the Legendre transform of $\gamma^v(k)$ that has unique minimum of zero at $\phi=\lambda_0$ where $\lambda_0\equiv d\gamma^v/dk(k=0)$. We conclude that
\begin{eqnarray}&&\!\!\!\!\!\!\!\!\!\!\!\!\!\!
\lim_{t\to\infty}\frac{\ln |\delta \bm v(t, \bm x)|}{t}=\lambda_0,
\end{eqnarray}
holds almost everywhere for almost every perturbed flow $\bm v(t, \bm x)$ and almost any initial perturbation field $\delta \bm v(t, \bm x)$. The above considerations indicate that it is plausible that $\lambda_0\tau_{\nu}$ is a weakly decaying function of $Re$. Strictly speaking the above considerations are a hypothesis which detailed study is left for future work.

\section{Conclusions}

In this work we established some properties of three sets of Lyapunov exponents associated with the Navier-Stokes turbulence. The first set $\gamma(k)$ describes the exponential divergence of two infinitesimally close Lagrangian trajectories of fluid particles. The second set $\gamma^B(k, \eta)$ is that of magnetic kinematic dynamo, the set which depends on the magnetic diffusivity $\eta$. Finally the third set $\gamma^v(k)$ provides infinite set of possible definitions of the Lyapunov exponent of the Navier-Stokes equations distinguished by the different definitions of the norm that is assigned to the flow perturbation field. It seems that this set has never been considered before, with the exception of the $L_2$ norm Lyapunov exponent.

We demonstrated using cumulant expansion theorem that $\gamma(k)$, defined by spatial averaging, is independent of the velocity realization and is given by a series whose terms are growing powers of $Re$. Using large deviations theory we demonstrated that $\gamma(k)$ plausibly obeys power-law behavior in $Re$ which agrees with all current numerical data. We demonstrated that the growth is hyper-intermittent.
A qualitative picture of the latter emerges by considering the flow regions where velocity gradients are anomalously large, exceeding the typical value of $\tau_{\nu}^{-1}$ by a power of $Re$. In those regions the stretching of an infinitesimal line element that connects the trajectories' positions occurs much faster than in typical regions with gradients of order $\tau_{\nu}^{-1}$.
%Moreover there are highly anomalous fluctuations  where these large gradient,
%PJK: added comma and plural for gradients
Moreover, there are highly anomalous fluctuations  where these large gradients,
%PJK
on (time) average, persist along the Lagrangian trajectory during an arbitrarily long time interval $t$. Since correlation time of these large gradients is finite then the probability of these persistent in time large fluctuations decays exponentially in time according to extremely rapid decay law $\sim \exp(-c t Re^{\alpha}/\tau_{\nu})$ with some constant $c$ and exponent $\alpha$. Still these events are not negligible because stretching of line elements on these events is extremely fast. The stretching rate is roughly constant and proportional to $Re^{\beta}$ with some exponent $\beta$ so that the distance grows proportionally to $\exp(c' t Re^{\beta}/\tau_{\nu})$ where $c'$ is a constant. As a result these events determine the growth rate of the moments since extreme smallness of the probability of the event is compensated by extreme largeness of averaged quantity on these events.

The $Re$ dependence of the exponents implies that care is needed in picking the numerical resolution of simulations to measure them, more so as higher $Re$ are considered. For the Lyapunov exponent of the fluid particles, which is decreased by intermittency in comparison with the Kolmogorov theory estimate, grid resolution with size of order of Kolmogorov scale will do since that is determined by the calm long-correlated turbulent events. In contrast, the Lyapunov exponent of the Navier-Stokes equations or the generalized Lyapunov exponent of the fluid particles, which are increased by intermittency in comparison with the Kolmogorov theory estimate, are determined by intermittent bursts whose scale is possibly much smaller than the Kolmogorov scale, cf. the study of dependence on lattice size in \cite{ho}.

The $Re-$dependence would hold also for many other similar quantities. This includes the counterpart of the generalized Lyapunov exponent for the vorticity stretching introduced in \cite{mj18}, the growth rate of moments of  concentration of inertial particles in turbulence \cite{arxiv}, the generalized Lyapunov exponent of magnetohydrodynamic turbulence \cite{ho} and others.

Theory of the Lyapunov exponent of the Navier-Stokes equations demands insight into the fastest growing perturbations. These were reported to be correlated with helicity \cite{hr} however the considered $Re$ are too small to make a conclusion on the infinite $Re$ limit. The reference studies the full spectrum of the Lyapunov exponents of the Navier-Stokes equations. It seems that $Re-$dependence of different exponents is different. This implies that there could be simplifications in the Kaplan-Yorke dimension of turbulence at $Re\to\infty$.

It is quite probable that there exists the generalization of multiplicative large deviations theory, as described by statistics of finite-time Lyapunov exponents to the case of infinite-dimensional operators such those that govern the growth of small perturbations of the Navier-Stokes equations or small fluctuations of magnetic field in turbulence of conducting fluids. The development of the corresponding formalism is a worthy research venue.

\section{ACKNOWLEDGMENTS}

We  would  like  to  acknowledge  the  support  of  Grant No. $2018118$ of  the U.S.-Israel  Binational  Science  Foundation.  This  research  was  supported  in  part  by  the Israel Science  Foundation(ISF)  under  Grant  No. 2040/17 and by  the Deutsche Forschungsgemeinschaft  Heisenberg programme (Grant No. KA$4825/2-1$).

%%%%%%%%%%%%%%%%%%%%%%%%%%%%%%%%%%%%%%%%%%%%%

%%%%%%%%%%%%%%%%%%%%%%%%%%%%%%%%%%%%%%%%%%%%%

\end{document}